\documentclass[floatfix,citeautoscript,nofootinbib,superscriptaddress,onecolumn,pra]{revtex4-2}
\usepackage{amsbsy}
\usepackage{latexsym,epsfig,graphicx}
\usepackage{dcolumn}
\usepackage{graphicx}
\usepackage{subfigure}
\usepackage{comment}
\usepackage{color}
\usepackage{bm}
\usepackage{mathrsfs}
\usepackage{amsfonts}
\usepackage{amsmath}
\usepackage{amssymb}
\usepackage{xspace}
\usepackage{epstopdf}
\usepackage{tabularx}
\usepackage{longtable}
\usepackage{wasysym}
\usepackage[colorlinks=true, letterpaper=true, pdfstartview=FitV, linkcolor=red, citecolor=blue, urlcolor=blue]{hyperref}
\usepackage[normalem]{ulem}

\pdfoutput=1
\begin{document}

\title{Dual opposing quadrature-PT symmetry}

\author{Wencong Wang}
\affiliation{Guangdong Provincial Engineering Research Center for Optoelectronic Instrument, School of Electronics and Information Engineering, South China Normal University, Foshan 528225, China}
\author{Jacob Kokinda}
\affiliation{Department of Electrical and Computer Engineering, North Carolina State University, Raleigh, North Carolina, 27695 USA}
\author{Jiazhen Li}
\affiliation{Department of Electrical and Computer Engineering, North Carolina State University, Raleigh, North Carolina, 27695 USA}
\author{Qing Gu}
\affiliation{Department of Electrical and Computer Engineering, North Carolina State University, Raleigh, North Carolina, 27695 USA}
\affiliation{Department of Physics, North Carolina State University, Raleigh, North Carolina, 27695 USA}
\author{Dongmei Liu}
\affiliation{Guangdong Provincial Engineering Research Center for Optoelectronic Instrument, School of Electronics and Information Engineering, South China Normal University, Foshan 528225, China}
\author{Jianming Wen}
\email{jianming.wen@kennesaw.edu}
\affiliation{Department of Physics, Kennesaw State University, Marietta, Georgia 30060, USA}
%% email address is required; see note below about the corresponding author designation

% use {asbstract*} to suppress the copyright line. Copyright information will be added in production

\begin{abstract} 
Our recent research on type-I quadrature parity-time (PT) symmetry, utilizing an open twin-beam system, not only enables observing genuine quantum photonic PT symmetry amid phase-sensitive amplification (PSA) and loss in the presence of Langevin noise but also reveals an additional classical-to-quantum (C2Q) transition in noise fluctuations. In contrast to the previous setup, our exploration of an alternative system assuming no loss involves a type-II PSA-only scheme. This scheme facilitates dual opposing quadrature PT symmetry, offering a comprehensive and complementary comprehension of C2Q transitions and anti-Hermiticity-enhanced quantum sensing. Furthermore, our investigation into the correlation with the Einstein-Podolsky-Rosen criteria uncovers previously unexplored connections between PT symmetry and nonclassicality, as well as quantum entanglement within the continuous-variable framework.

\end{abstract}

\maketitle
%%%%%%%%%%%%%%%%%%%%%%%%%%  body  %%%%%%%%%%%%%%%%%%%%%%%%%%
\section{Introduction}
Over the past decade, classical linear and nonlinear photonic systems, characterized by gain and loss, have served as a robust and versatile platform for exploring non-Hermitian (NH) physics. Notably, these systems have played a key role in probing parity-time (PT) symmetry \cite{1,2,3,4,5,6,7,8,9,10,11,12,13,14,15,16,17}, unveiling a variety of peculiar effects absent in Hermitian counterparts. These effects include the typical PT phase transition, where eigenvalues transition from real to imaginary, and the coalescence of eigenvalues and eigenvectors at the phase transition point, known as the exceptional point (EP). The significant achievements in classical PT systems have prompted a recent shift in focus towards open quantum optical systems \cite{18,19,20,21,22,23,24,25} with an aim to disclose distinctive quantum features. 

However, challenges \cite{26,27} like Langevin noise, the quantum noncloning theorem, and the causality principle have led to the prevailing belief that quantum optical PT symmetry with both gain and loss is unlikely. This limitation confines research mainly to dissipative single-partite systems--such as single photons \cite{18,22}, ultracold atoms \cite{21}, trapped ions \cite{19,23}, nitrogen-vacancy centers \cite{20} in diamond, and superconducting qubits \cite{24}--utilizing postselection measurement. Consequently, observed PT phase transitions closely resemble classical NH scenarios, rooted in a (semi)classical interpretation. Unfortunately, these studies fail to offer insights on the viability of gain-loss-coupled quantum optical PT symmetry or unique effects exclusive to quantum NH settings absent in classical NH or Hermitian quantum counterparts.

Unlike prior research, our recent work \cite{28} demonstrates the attainability of genuine quantum optical PT symmetry in an open twin-beam system through four-wave mixing (FWM), overcoming these challenges by employing phase-sensitive amplification (PSA) instead of phase-insensitive amplification (PIA) and studying field quadrature observables. Remarkably, under fair sampling measurement, our PSA-loss bipartite system establishes unique type-I quadrature PT symmetry without a classical analog, introducing an additional dynamical or stationary classical-to-quantum (C2Q) transition in quadrature noise fluctuations alongside the standard PT phase transition in eigenvalues. The emergence of these dual transitions in the continuous-variable (CV) framework is a minimum signature for claiming quantum behavior, as further supported by our recent studies \cite{29,30} on a dissipative spin-boson-coupled superconducting circuit platform, showcasing the co-emergence of eigenspectral phase transition and exceptional entanglement transition in the Fock space through post-projection measure.

\begin{figure}[htbp]
\centering\includegraphics[width=1\linewidth]{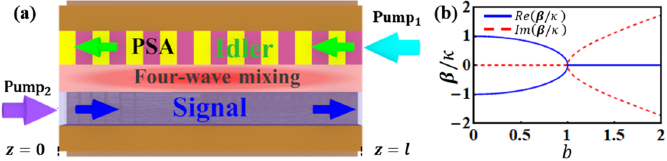}
\caption{(a) Dual opposing type-II quadrature PT symmetry emerges in twin-beam generation, with the backward-propagating idler mode undergoing PSA at a rate of $g$, while the forward-propagating signal experiences lossless transmission. (b) Regular PT phase transition associated with eigenvalues $\pm\beta$.}
\label{fig:scheme}
\end{figure}

In our previous vacuum-input type-I quadrature-PT configuration \cite{28}, the system displays unconventional features challenging conventional expectations of quantum squeezing. Notably, there is no need for a cavity or high parametric gain, and the system exhibits anomalous loss-induced quadrature squeezing. Moreover, while the quadrature PT behaviors manifest various sharp C2Q transitions related to quadrature noise fluctuations, the EP generally does not coincide with these C2Q transition boundaries. Furthermore, the nontrivial interference between PSA or loss and parametric conversion expedites the emergence of nonclassical correlations beyond traditional quantum squeezing, necessitating additional conditions. Here, we deepen our understanding of quadrature PT symmetry and its effects in the twin-beam system by considering no loss and only involving PSA (Fig.~\ref{fig:scheme}(a)). Differing from the previous PSA-loss (type-I) case, both quadrature pairs in this type-II PSA-only scheme evolve with contrasting PT symmetry, resulting in dual opposing quadrature PT symmetry. Additionally, this type-II system emerges as a quintessential platform to unveil various intriguing physics including its profound connection with the renowned Einstein-Podolsky-Rosen (EPR) correlations, a territory largely uncharted in NH settings thus far.

\section{Theoretical model}
Assuming undepleted and classical input pump lasers, the evolution of correlated signal-idler field operators $a_s$ and $a_i$ along the $\pm z$-direction is governed by the Hamiltonian $H=i\hbar g(a^2_i-a^{\dagger 2}_i)/2+\hbar\kappa(a^{\dagger}_ia^{\dagger}_s+a_ia_s)$. The corresponding Heisenberg equations are $da_i/dz=ga^{\dagger}_i+i\kappa a^{\dagger}_s$ and $da_s/dz=-i\kappa a^{\dagger}_i$, where $\dagger,g,\kappa$ denote the Hermitian conjugate, PSA, and FWM parametric conversion coefficient, respectively. Akin to our latest work \cite{28}, hidden PT symmetry emerges upon transforming them into coupled-quadrature forms,
\begin{subequations}
\begin{align}
\frac{d}{dz}\begin{bmatrix}
q_i\\
p_s
\end{bmatrix}&=\begin{bmatrix}
g & \kappa\\
-\kappa & 0
\end{bmatrix}\begin{bmatrix}
q_i\\
p_s
\end{bmatrix}, \label{eq:quadrature1} \\
\frac{d}{dz}\begin{bmatrix}
p_i\\
q_s
\end{bmatrix}&=\begin{bmatrix}
-g & \kappa\\
-\kappa & 0
\end{bmatrix}\begin{bmatrix}
p_i\\
q_s
\end{bmatrix}, \label{eq:quadrature2} 
\end{align}
\end{subequations}
by introducing $q_j=(a^{\dagger}_j+a_j)/2$ and $p_j=i(a^{\dagger}_j-a_j)/2$ ($j=i,s$) with the commutation relation $[q_j,p_j]=i/2$. As detailed in the Supplementary Information (SI), the quadrature pair $\{q_i,p_s\}$ follows active PT symmetry, while the other pair $\{p_i,q_s\}$ obeys passive PT symmetry after opposite gauge transformations. Both pairs share the same EP at $b=g/2\kappa=1$, with an identical pair of eigen-propagation values $\pm\beta=\pm\kappa\sqrt{1-b^2}$ phase transitioning from real to imaginary for $b<1$ and $b>1$. This striking phenomenon, termed \textit{dual opposing quadrature-PT symmetry}, represents a pure quantum effect which is inaccessible in PIA-based structures. For a medium of length $l$, the solutions of Eqs.~(\ref{eq:quadrature1}) and (\ref{eq:quadrature2}) are
\begin{subequations}
\begin{align}
\begin{bmatrix}
q_i(0)\\
p_s(l)
\end{bmatrix}&=\frac{-\sin\epsilon}{\sin(\beta l+\epsilon)}\begin{bmatrix}
-e^{-\frac{gl}{2}} & \frac{\sin(\beta l)}{\sin\epsilon}\\
\frac{\sin(\beta l)}{\sin\epsilon} & -e^{\frac{gl}{2}}
\end{bmatrix}\begin{bmatrix}
q_i(l)\\
p_s(0)
\end{bmatrix}, \label{eq:solution1} \\
\begin{bmatrix}
q_s(l)\\
p_i(0)
\end{bmatrix}&=\frac{\sin\epsilon}{\sin(\beta l-\epsilon)}\begin{bmatrix}
-e^{-\frac{gl}{2}} & \frac{\sin(\beta l)}{\sin\epsilon}\\
\frac{\sin(\beta l)}{\sin\epsilon} & -e^{\frac{gl}{2}}
\end{bmatrix}\begin{bmatrix}
q_s(0)\\
p_i(l)
\end{bmatrix}, \label{eq:solution2} 
\end{align}
\end{subequations}
where $\epsilon=\tan^{-1}(2\beta/g)$ is a PT-induced phase shift. Their physics significance will become clear shortly. 

\section{Homodyne detection}
\begin{figure}[htbp]
\centering\includegraphics[width=1\linewidth]{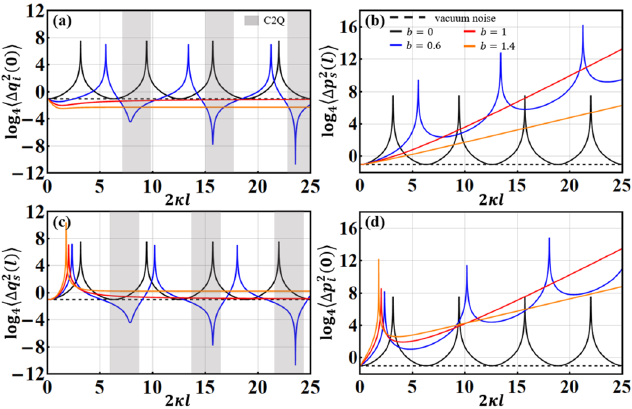}
\caption{Comparing with standard TMSV ($b=0$, black solid line) and vacuum noise (black dashed line), the quadrature variances $\{\langle\Delta q^2_i(0)\rangle,\langle\Delta p^2_s(l)\rangle\}$ (a,b) and $\{\langle\Delta q^2_s(l)\rangle,\langle\Delta p^2_i(0)\rangle\}$ (c,d) feature active and passive PT-induced quadrature squeezing and de-squeezing, respectively. In the PT phase-unbroken region ($b<1$), $\langle\Delta q^2_i(0)\rangle$ and $\langle\Delta q^2_s(l)\rangle$ concurrently display flexible-length dynamical C2Q transitions; whereas complementary stationary C2Q transitions occur for a fixed length (shaded grey areas) with the EP-variance curves serving as the exact boundary between the incompatible classical and quantum noise realms.}
\label{fig:singlevariance}
\end{figure}

The concise and symmetrical structures of Eqs.~(\ref{eq:solution1}) and (\ref{eq:solution2}) render them well-suited and mutually complementary for probing the fundamental nature of dual opposing quadrature PT symmetry. They offer valuable insights into physical phenomena, including versatile C2Q transitions, and effectively address causality concerns, particularly in single-mode quadrature scenarios. We thus focus on four single-mode quadrature variances, $\langle\Delta q_j^2\rangle=\langle q^2_j\rangle-\langle q_j\rangle^2$ and $\langle\Delta p^2_j\rangle=\langle p^2_j\rangle-\langle p_j\rangle^2$, leaving the two-mode quadrature and relative-intensity squeezing cases for the SI. Applying Eqs.~(\ref{eq:solution1}) and (\ref{eq:solution2}), we arrive at exact expressions for these four variances,
\begin{subequations}
\begin{align}
\langle\Delta q^2_i(0)\rangle&=\frac{e^{-gl}\sin^2\epsilon+\sin^2(\beta l)}{4\sin^2(\beta l+\epsilon)}, \label{eq:qivariance} \\
\langle\Delta p^2_s(l)\rangle&=\frac{e^{gl}\sin^2\epsilon+\sin^2(\beta l)}{4\sin^2(\beta l+\epsilon)}, \label{eq:psvariance}\\
\langle\Delta q^2_s(l)\rangle&=\frac{e^{-gl}\sin^2\epsilon+\sin^2(\beta l)}{4\sin^2(\beta l-\epsilon)}, \label{eq:qsvariance} \\
\langle\Delta p^2_i(0)\rangle&=\frac{e^{gl}\sin^2\epsilon+\sin^2(\beta l)}{4\sin^2(\beta l-\epsilon)}, \label{eq:pivariance}
\end{align}
\end{subequations}
where Eqs.~(\ref{eq:qivariance}) and (\ref{eq:psvariance}) are mathematically symmetric to (\ref{eq:qsvariance}) and (\ref{eq:pivariance}). For active PT-symmetric $\{q_i(0),p_s(l)\}$, superluminal (fast) light effects are expected, indicated by an advanced phase shift $+\epsilon$ in the variances. Conversely, passive PT-symmetric $\{q_s(l),p_i(0)\}$ are anticipated to manifest subluminal (slow) light effects, evident in a phase delay $-\epsilon$ in their variances. Causality is maintained despite potential effects, as individual signal or idler fields exist in mixed states. In the presence of optical loss \cite{28}, Langevin noise obscures the observation of slow- and fast-light effects, making it challenging to detect them in the type-I scenario. Besides, PT symmetry disrupts symmetric noise characteristics and $2\pi$-periodicity in usual two-mode squeezed vacuum (TMSV), facilitating the rapid emergence of quantum squeezing and exceptional C2Q transitions. The implications of these results are apparent in the numerical plots against the dimensionless propagation variable $2\kappa l$ for various $b$ during $\pm\beta$-phase transitions (Figs.~\ref{fig:singlevariance}(a)-(d)). Under the same $\kappa$, the ideal TMSV ($g=0$, solid black curves) fails to yield quantum squeezing, with each quadrature variance oscillating above the vacuum noise (dashed black lines). 

For active quadrature PT symmetry in its phase-unbroken region ($b<1$), logarithmic $\langle\Delta q^2_i(0)\rangle$ (Fig.~\ref{fig:singlevariance}(a)) periodically fluctuates at a new period $T=2\pi\kappa/\beta$. Gradually growing squeezing peaks at trough locations $nT$ (with $n$ as positive integers), even without a cavity, signifying flexible-length dynamical C2Q transitions with other fixed parameters. In the phase-broken regime ($b>1$), periodic noise distributions cease, and quantum noise reduction remains consistently available. A larger $b$ results in greater quantum squeezing. Importantly, the variance curve at the EP serves as the partition line, distinguishing these two distinct noise behaviors. Note that the stationary C2Q transition (shaded gray areas) can also occur at a fixed length by changing only $g$, where the EP-variance again acts as the exact boundary dividing the incompatible classical and quantum noise worlds. Contrastingly, logarithmic $\langle\Delta p^2_s(l)\rangle$ (Fig.~\ref{fig:singlevariance}(b)) assumes similar periodic classical fluctuations peaking at $(nT-2\epsilon\kappa/\beta)$ for the intact PT phase. As symmetry spontaneously breaks down, it grows monotonically with an upper bound set by the EP-variance, and the greater the $b$ value, the less the de-squeezing. 

For passive quadrature-PT, $\langle\Delta q^2_s(l)\rangle$ and $\langle\Delta p^2_i(0)\rangle$ (Figs.~\ref{fig:singlevariance}(c) and (d)) behave similarly despite with distinct patterns. Specifically, when $b<1$, $\langle\Delta q^2_s(l)\rangle$ shows noticeable quadrature squeezing, akin to $\langle\Delta q^2_i(0)\rangle$, but with periodic squeezing amplitudes shifting to $[(n-1)T+2\epsilon\kappa/\beta]$, implying flexible-length dynamical C2Q transitions. In contrast, $\langle\Delta p^2_i(0)\rangle$ amplifies classical noise with periodic fluctuations, similar to $\langle\Delta p^2_s(l)\rangle$. When $b\geqslant1$, apart from a lone peak at $l=\epsilon/\beta$, $\langle\Delta q^2_s(l)\rangle$ is anti-squeezed, and increasing $b$ intensifies the anti-squeezing. Compared to $\langle\Delta q^2_i(0)\rangle$, a complementary stationary C2Q transition at a fixed length (shaded grey areas) appears alongside the $\pm\beta$-based PT-phase transition by manipulating only $g$. For $\langle\Delta p^2_i(0)\rangle$, it exhibits classical noise amplification regardless of PT symmetry collapse, akin to $\langle\Delta p^2_s(l)\rangle$. The distinction between these variances is evident. The periodic fluctuations of $\langle\Delta p^2_i(0)\rangle$ (Fig.~\ref{fig:singlevariance}(d)) precede those of $\langle\Delta p^2_s(l)\rangle$ by a net phase $2\kappa(2\epsilon-\pi)/\beta$ before the PT-phase breaks. After the phase breaks, $\langle\Delta p^2_i(0)\rangle$ always peaks around $l=\epsilon/\beta$ before growing monotonically. The separation distance between any two adjacent peaks (or valleys) is $(\pi-2\epsilon)/\beta$, uniform for all four periodically fluctuating variances. Previous work \cite{28} has documented that the presence of a balanced loss to the signal mode diminishes the symmetry of the entire system, inducing divergent courses in the two sets of conjugate quadrature pairs. 

\section{Nonclassicality and EPR Correlation}
Other than showing the dual transitions, this succinct system is an excellent toolbox for disclosing the singular relation between PT symmetry and nonclassicality, an emerging frontier barely touched to date. To delve into this, we first employ the EPR criteria proposed by Reid \cite{31,32} to seek resolutions. We note that the four single-mode quadratures~(\ref{eq:solution1}) and (\ref{eq:solution2}) coincide with their respective amplitudes and phases, namely, $X_1(0)=q_i(0), X_2(0)=p_i(0), Y_1(l)=q_s(l)$, and $Y_2(l)=p_s(l)$ (SI). Consequently, the Cauchy-Schwarz inequality for these quadrature phase amplitudes can be explored by examining the quantum-mechanical correlation coefficients,
%\begin{equation}
%    |\langle q_i(0)q_s(l)\rangle|^2\leqslant\langle[q_i(0)]^2\rangle\langle[q_s(l)]^2\rangle.\nonumber
%\end{equation}
\begin{equation}
    |C_{jm}|=\left\vert\frac{\langle X_j(0)Y_m(l)\rangle}{\sqrt{\langle[X_j(0)]^2\rangle\langle[Y_m(l)]^2\rangle}}\right\vert\leqslant1, \label{eq:Cdefination}
\end{equation}
with $\{j,m\}=\{1,2\}$. After some calculations (SI), we have
\begin{subequations}
\begin{align} 
C_{jj}&=0,\label{eq:c1}\\
C_{jm}&=\frac{2\cosh{(gl/2)}\sin\epsilon\sin(\beta l)}{\sqrt{\sin^4\epsilon+\sin^4(\beta l)+2\cosh(gl)\sin^2\epsilon\sin^2(\beta l)}},\label{eq:c2}
\end{align}
\end{subequations}
with $j\neq m$. Physically, $|C_{jm}|=1$ means a state of perfect quantum correlation between $X_j(0)$ and $Y_m(l)$, whereas $|C_{jm}|=0$ implies a complete absence of such correlation between the two. 

\begin{figure}[htbp]
\centering\includegraphics[width=1\linewidth]{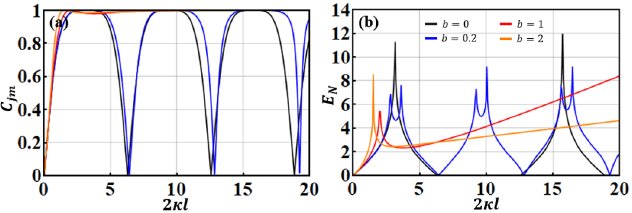}
\caption{Assessing PT symmetry's influence on nonclassicality using quantum correlation coefficient $C_{jm}$ (a) and logarithmic negativity $E_N$ (b), in comparison to the standard TMSV case.}
\label{fig:entanglement}
\end{figure}

Therefore, Eq.~(\ref{eq:c1}) denotes a complete lack of nonclassical correlation between $X_1(0)$ ($X_2(0)$) and $Y_1(l)$ ($Y_2(l)$) regardless of the presence of PT symmetry. In contrast, the situation described by Eq.~(\ref{eq:c2}) becomes subtle for the pair $\{X_1(0),Y_2(l)\}$ (or $\{X_2(0),Y_1(l)\}$). To see its behavior, Fig.~\ref{fig:entanglement}(a) numerically plots $C_{12}$ (or $C_{21}$), revealing abrupt changes made by PT symmetry. Prior to the phase breaking ($b<1$), $|C_{12}|$ or $|C_{21}|$ oscillates between 1 and 0 at the period $2\pi\kappa/\beta$, but differs substantially from the conventional case ($g=0$, black line), demonstrating periodic shifts between the perfect existence and complete absence of quantum correlation. Contrarily, after the phase transition ($b>1$), it asymptotically approaches unity, implying a perfect quantum correlation between the cross quadrature phase amplitudes. Here the EP-line simply acts as the circumscription to differentiate curve patterns across the PT phase transition. In essence, the scenario resembles the EPR \textit{Gedankenexperiment} \cite{33}. In the $b\geqslant1$ region, maximum correlation between amplitude $X_1(0)$ and $Y_2(l)$, as well as between $X_2(0)$ and $Y_1(l)$, can be easily achieved, leading to the application of EPR reasoning. In the $b<1$ regime, the EPR scenario applies broadly to these two quadrature phase amplitude pairs, although it may depend on the propagation distance. It is worth noting that the influence of anti-Hermiticity on EPR correlations here is analogous to the impact of non-Hermiticity on quantum entanglement in a dissipative spin-boson system--that is, PT symmetry gives rise to extra exceptional nonclassical phenomena \cite{29,30}.

The extent of nonclassicality in the system can be further assessed through the analysis of logarithmic negativity \cite{34,35}, denoted as $E_N=\textrm{max}[0,-\ln{4\eta}]$, derived from the system's 4 by 4 covariance matrix $V_Q=[A,C;C^{\mathrm{Tr}},B]$ (see SI). Here, $\eta=\sqrt{(\Sigma-\sqrt{\Sigma^2-4\mathrm{det}V_Q})/2}$ and $\Sigma=\mathrm{det}A+\mathrm{det}B-2\mathrm{det}C$ with $A=[\langle\Delta q^2_i(0)\rangle,0;0,\langle\Delta p^2_i(0)\rangle]$, $B=[\langle\Delta q^2_s(l)\rangle,0;0,\langle\Delta p^2_s(l)\rangle]$, $C=[0,\langle q_i(0)p_s(l)\rangle-\langle q_i(0)\rangle\langle p_s(l)\rangle;\langle q_s(l)p_i(0)\rangle-\langle q_s(l)\rangle\langle p_i(0)\rangle,0]$, and Tr representing transpose. The farther the quantum correlation that $E_N$ reflects, the farther it is from 0. In Fig.~\ref{fig:entanglement}(b), we present representative examples both before and after the $\beta$-phase transition, comparing them with the TMSV case (black). For $b<1$, $E_N$ exhibits gradually increasing bimodal cyclical oscillations above zero, reaching zero only at valleys. This signifies a substantially enlarged range with quantum emergence. In contrast, for $b\geqslant1$, $E_N$ develops a single peak that is always larger than 0, indicating a full spectrum of quantum availability. As evident, the analysis of $E_N$ aligns with the previously discussed analysis of $C_{jm}$ $(j\neq k)$ except from the entire system perspective.

\section{Quantum sensing}
As a pivotal nonclassical resource, quantum squeezing \cite{36} plays a key role in traditional quantum sensing and metrological applications \cite{37}. Recent studies \cite{38,39} leveraging anti-PT symmetry have demonstrated enhanced squeezing-based quantum sensitivity near EP. This prompts an exploration of whether improved sensitivity is attainable in our type-II quadrature-PT system. In contrast to the type-I quadrature-PT case \cite{28}, the type-II system arises as a versatile PT-enhanced quantum sensor with unparalleled performance, surpassing designs based on squeezing factors or EP alone. Besides, passive PT quadrature outperforms active PT quadrature, and two-mode quadrature outshines single-mode quadrature (SI) under the same system parameters. The achievable precision approaches the quantum Cram\'{e}r-Rao bound, dictated by the quantum Fisher information (QFI) of the quantum state, although it experiences a loss of sensitivity near and above the EP. 

To commence, we assume the initial preparation of the two bosonic modes in a two-photon coherent state, $|\Phi\rangle_0=|\alpha_i,\alpha_s\rangle$. Subsequently, we set $\alpha_i=i\alpha^*_s\equiv\sqrt{2}\alpha e^{i\pi/4}$ for simplification in the upcoming calculations. In the case of single-mode quadrature, homodyning detection is applied at an interaction distance $l$ to an observable, say, $q_i(0)$. Utilizing Eq.~(\ref{eq:solution1}), we find the mean value and variance of $q_i(0)$ to be
\begin{eqnarray}
\langle q_i(0)\rangle =\frac{e^{- \frac{gl}{2}}\sin\epsilon\langle q_i(l)\rangle-\sin(\beta l)\langle p_s(0)\rangle}{\sin(\beta l+\epsilon)}.\label{eq:qexp}
\end{eqnarray}
The ultimate accuracy of sensing relies on the precision with which a small change in $\langle q_i(0)\rangle$ can be measured in response to a tiny perturbation $\delta\kappa$ from a predefined $\kappa$-value. This system response is characterized by a susceptibility (SI), denoted as $\chi^{q_i(0)}_{\kappa}$, defined as $\partial\langle q_i(0)\rangle/\partial\kappa$ and derived from Eq.~(\ref{eq:qexp}). Next, the achievable accuracy of estimating the parameter $\kappa$'s precision can be evaluated by considering the variance (\ref{eq:qivariance}) and susceptibility. This evaluation involves the relation $\Delta\kappa^2_{q_i(0)}=\langle\Delta q^2_i(0)\rangle/[\chi^{q_i(0)}_{\kappa}]^2$. 

\begin{figure}[htbp]
\centering\includegraphics[width=1\linewidth]{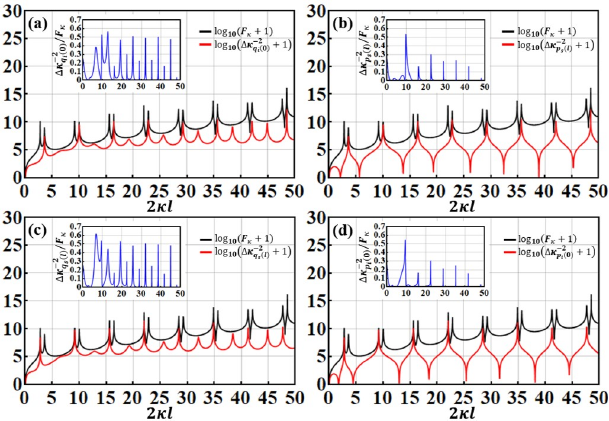}
\caption{Quadrature PT-symmetric quantum sensing. The ratios of the inverse variances $\log_{10}({\Delta\kappa}_{q_i(0)}^{- 2}+1)$ (a), $\log_{10}({\Delta\kappa}_{p_s(l)}^{- 2}+1)$ (b), $\log_{10}({\Delta\kappa}_{q_s(l)}^{- 2}+1)$ (c), and $\log_{10}({\Delta\kappa}_{p_i(0)}^{- 2}+1)$ (d) of the four observables to the quantum Fisher information $\log_{10}(F_{\kappa}+1)$ as functions of dimensionless lengths for parameters $\{\alpha=10,\kappa=0.5\}$, illustrating the sensitivity enhancement in the quadrature-PT phase unbroken region ($b=0.2$).}
\label{fig:sensing}
\end{figure}

The inverse variance $\Delta\kappa^{-2}_{q_i(0)}$ determines the sensing power of the system, with its upper bound constrained by the QFI, $F_{\kappa}$. The QFI establishes the lower quantum Cram\'{e}r-Rao bound, expressed as $F_{\kappa}\geqslant\Delta\kappa^{-2}_{q_i(0)}$, representing the utmost precision attainable through optimal measurement. For the seeding coherent input state $|\Phi\rangle_0$, the QFI of the system can be deduced from (see SI for more detail)
\begin{equation}
F_{\kappa}=\left( \frac{d\mu}{d\kappa} \right)^{\intercal}V^{- 1}\frac{d\mu}{d\kappa}. \label{eq:fk3}
\end{equation}
with $\intercal$ for a vector or matrix transpose. In Eq.~(\ref{eq:fk3}), the amplitude vector $\mu$ and $V^{-1}$ are computed in the quadrature basis via $\mu_j=\langle{\hat{v}}_{j}\rangle$ and $V_{j,k} = \frac{1}{2}\left\langle {{\hat{v}}_{j}{\hat{v}}_{k} + {\hat{v}}_{k}{\hat{v}}_{j}} \right\rangle - \left\langle {\hat{v}}_{j} \right\rangle\left\langle {\hat{v}}_{k} \right\rangle$, for $1 \leq j,k \leq 2$, with the column vector $\hat{v} = \left( {q_{i}(0),q_{s}(l),p_{i}(0),p_{s}(l)} \right)^{\intercal}$. As $F_{\kappa}$ ascribes the entire system, it aligns with all other quantum observables. Notably, $F_{\kappa}$ manifests differently in response to the contrasting PT domains. In Fig.~S7, we present logarithmic $F_{\kappa}$ with distinctive characteristics in three circumstances: $b=0.2$ (unbroken PT phase), $b=1$ (EP), and $b=2$ (breaking PT phase). Similarly, we can test $\kappa$-parameter estimation using the rest three single-mode quadratures (SI): $p_i(0), q_s(l)$, and $p_s(l)$. Our calculations indicate that our current scheme optimally supports quantum sensor performance in the quadrature-PT phase unbroken regime but away from the EP, distinguishing it from EP-based sensors. Numerical simulations in Fig.~\ref{fig:sensing} demonstrate that the best sensing is achieved with a suitable medium length $l$, owing to the enlarged Hilbert space of the final system state. We illustrate four inverse variances before the quadrature-PT phase transitions in Figs.~\ref{fig:sensing}(a)-(d), portraying the ratios (insets) of $\Delta\kappa^{-2}_{q_i(0)}$, $\Delta\kappa^{-2}_{p_i(0)}$, $\Delta\kappa^{-2}_{q_s(l)}$ and $\Delta\kappa^{-2}_{p_s(l)}$ to $F_{\kappa}$, providing an instructive view of the data. Upon careful scrutiny of Fig.~\ref{fig:sensing}, we deduce that both $q_i(0)$ and $q_s(l)$ supply the maximum parameter estimation precision at the first oscillating peaks of $\log_{10}({\Delta\kappa}_{q_i(0)}^{- 2}+1)$ and $\log_{10}({\Delta\kappa}_{p_s(l)}^{- 2}+1)$ with respect to $\log_{10}(F_{\kappa}+1)$. However, overall, $q_s(l)$ outperforms $q_i(0)$ in sensing performance in terms of sensitivity and accuracy, entailing that the variance $\Delta\kappa^2_{q_s(l)}$ of passive quadrature-PT symmetry is smaller than the variance $\Delta\kappa^2_{q_i(0)}$ of active quadrature-PT. As $l$ increases, the sensing ability of both observables worsens because $F_{\kappa}$ gets larger, and the ratios become smaller. By contrast, $p_i(0)$ and $p_s(l)$ allow classical sensing exclusively with non-equivalent performance.

\section{Conclusion}
In contrast to previous studies, the type-II PSA-only twin-beam system serves as an optimal quantum platform for investigating dual opposing quadrature-PT symmetry. Our sophisticated theoretical framework allows for the exploration of profound C2Q transitions and establishes a unique link between PT symmetry, nonclassicality, and the EPR paradox. Our findings unveil nontrivial quantum aspects of PT symmetry beyond the capabilities of non-PSA-based systems, demonstrating promise for designing PT-enhanced quantum sensors that leverage anti-Hermiticity and squeezing. We anticipate that our proposed quadrature-PT scheme will unlock novel physical phenomena, enabling nonreciprocal transmission of continuous-variable (CV) qubits with minimal noise--crucial for CV-based quantum information processing and computing \cite{40}. This capability poses a challenge for PIA-based apparatus \cite{4,5,6,7,8,9,10,11,12,13,14,15,16,17,25,26,27}. By incorporating PSA, our system protects CVs from decoherence, compensating for loss through noiseless amplification and advancing CV-based quantum technologies. The nontrivial modulation and acceleration of quantum squeezing generation by PT symmetry further contribute to the development of diverse CV-based quantum technologies.

\section*{Author contributions} J.W. conceptualized the idea and oversaw the entire project, receiving support from Q.G. and D.L. W.W. conducted the theoretical calculations with assistance from J.K. and J.L., under the supervision of J.W. All authors contributed to the discussions and writing of the manuscript.

\section*{Funding} This work was supported by NSF ExpandQISE-2329027. J. Kokinda, J.Li, and Q.Gu also acknowledge support from NSF ECCS-2240448. W. Wang and D. Liu were supported by the National Key R$\&$D Program of China (2021YFA1400803) and GuangDong Basic and Applied Basic Research Foundation (2022A1515140139).

\section*{Acknowledgments} We thank S.-W. Huang for enlightening discussions.

\section*{Disclosures} The authors declare no conflicts of interest.

\section*{Supplemental document}
See Supplement 1 for supporting content. 

% Bibliography

\end{document}

\title{Supplementary Materials for Dual opposing quadrature-PT symmetry}

\begin{abstract}
\end{abstract}

\maketitle

\section{Further discussion on type-II quadrature PT symmetry}
In our previous work \cite{1}, we conducted theoretical research on type-I quadrature PT symmetry within a backward nonlinear parametric optical process coupled with phase-sensitive linear quantum amplification (PSA) featuring balanced loss. We discovered that in type-I scenarios, only one pair of quadratures exhibits PT symmetry with intriguing dual transitions—namely, a conventional PT phase transition linked to eigenvalues and a unique classical-to-quantum (C2Q) transition in quadrature noise fluctuations—while the other pair displays anomalous loss-induced quadrature squeezing. This disparity between the two quadrature pairs piqued our curiosity about the outcome in the absence of loss within the process, leading us to investigate type-II quadrature PT symmetry in our current work. Due to the straightforward nature of the type-II configuration, it enables us to carry out analytical investigations into the underlying physics of various topics that might pose challenges in a type-I configuration.

\subsection{Further remarks on Equations (1a) and (1b) in the main text}
Here, we would like to provide additional remarks regarding Eqs. (1a) and (1b) in the main text. In the derivation of these two sets of coupled quadrature equations, we assumed a squeezing angle of $\theta=0^{\circ}$ in the general quadrature operators, specifically $X_{1j}=({a_{j}^{\dagger}e^{i\theta} + a_{j}e^{- i\theta}})/2$ and $X_{2j} = i( {a_{j}^{\dagger}e^{i\theta}-a_{j}e^{- i\theta}})/2$. Additionally, from these equations, we derived the effective NH Hamiltonian matrices governing the dynamics of the quadrature pairs $\left\{ {q_{i},p_{s}} \right\}$ and $\left\{ {p_{i},q_{s}} \right\}$, which establish the foundation of our quaduature PT symmetry, taking the form:
\begin{subequations}
\begin{align}
H_{({q_{i},p_{s}})} &= \begin{bmatrix}
{ig} & {i\kappa} \\
{- i\kappa} & 0
\end{bmatrix} = i\frac{g}{2}\mathbb{I} + \begin{bmatrix}
{i\frac{g}{2}} & {i\kappa} \\
{- i\kappa} & {- i\frac{g}{2}}
\end{bmatrix} = i\frac{g}{2}\mathbb{I} + H_{a},\label{eq:H_qips}\\
H_{({p_{i},q_{s}})} &= \begin{bmatrix}
{- ig} & {i\kappa} \\
{- i\kappa} & 0
\end{bmatrix} = - i\frac{g}{2}\mathbb{I} + \begin{bmatrix}
{- i\frac{g}{2}} & {i\kappa} \\
{- i\kappa} & {i\frac{g}{2}}
\end{bmatrix} = - i\frac{g}{2}\mathbb{I} + H_{p}.\label{eq:H_piqs}
\end{align}
\end{subequations}
Here, $\mathbb{I}$ denotes the 2×2 unit matrix, and $H_{a} + H_{p}^{\top} = 0$, where $\top$ stands for the transpose of a matrix. It is evident that both quadrature pairs evolve PT-symmetrically, satisfying $\left\lbrack {H_{({q_{i},p_{s}})},PT} \right\rbrack = \left\lbrack {H_{({p_{i},q_{s}})},PT} \right\rbrack = 0$, where P represents the parity operator $\begin{bmatrix}
0 & 1 \\
1 & 0
\end{bmatrix}$ and $T$ assumes the complex conjugation. Of importance, according to Eqs.~(\ref{eq:H_qips}) and (\ref{eq:H_piqs}), it is immediately apparent that $\left\{ {q_{i},p_{s}} \right\}$ exhibits active PT symmetry while $\left\{ {p_{i},q_{s}} \right\}$ demonstrates passive PT symmetry. Both pairs share the same EP at $\kappa = g/2$ (i.e., $b=1$) and the same pair of eigen-propagation constants ($\pm \beta = \pm ~\kappa\sqrt{1 - b^{2}}$), transiting from purely real ($b<1$) to conjugate imaginary ($b>1$). We are ware that Eqs.~(1a) and (1b) in the main text can be directly derived by recasting the given Hamiltonian in the quadrature format, $H=\hbar g(q_ip_i+p_iq_i)-2\hbar\kappa(q_iq_s-p_ip_s)$.

We note previous exploration of the coexistence of dual active-passive PT symmetry in the classical regime, achieved through coupled microcavities and the directional nature of parametric gain driven by momentum conservation \cite{2}. This leads to an intriguing phenomenon where an input signal traveling in one direction experiences parametric gain and adheres to active PT symmetry, while its counterpart traveling in the opposite direction remains unaffected by the gain, exhibiting passive PT symmetry. This yields what is known as \textit{non-reciprocal PT symmetry} for signal field launching in different directions \cite{2}. 

However, our type-II quadrature PT displays a fundamentally different behavior from this classical scenario. Here, the active and passive PT symmetry experienced by the two sets of conjugate quadrature pairs pertain to the same light field. Furthermore, in addition to the eigenvalues-based PT phase transition, the dynamic and stationary C2Q transitions in noise fluctuations also occur in these two sets of quadrature pairs, but in a complementary manner, as demonstrated in the single-mode quadrature case in the main text. 

In the subsequent discussion, we will illustrate that, similar to type-I quadrature PT symmetry \cite{1}, these dual transitions, especially the C2Q transitions, can also be observed in two-mode quadrature homodyne detection and relative-intensity squeezing measurement (RISM), albeit with distinct characteristics.

\subsection{Two-mode homodyne detection}
As shown in the main text, the PT-inherent eigenvalues $\pm\beta$ modulate the corresponding single-mode quadrature properties through noise fluctuations, resulting in both dynamic and stationary C2Q transitions. Our earlier investigation on type-I quadrature PT symmetry \cite{1} revealed that these C2Q transitions can also be observed in two-mode quadrature homodyning measurements. Therefore, it is interesting to explore the behaviors of these C2Q transitions by measuring two-mode quadratures in the present type-II scenario. 

Similar to type-I, we start with the following two-mode quadratures, $d_1=[q_i(0)+q_s(l)]/\sqrt{2}$ and $d_2=[p_i(0)+p_s(l)]/\sqrt{2}$, which are linear combinations of the signal-idler field quadrature operators and satisfy the commutation relation $[d_1,d_2]=i/2$. For the vacuum input state, the variances of $d_1$ and $d_2$ are simply the sum of the corresponding single-mode quadrature variances (3a)--(3d) given in the main text,
\begin{subequations}
\begin{align}
\langle\Delta d^2_1\rangle&= \left\langle \left( \frac{q_{i}(0) + q_{s}(l)}{\sqrt{2}} \right)^{2} \right\rangle - \left\langle \frac{q_{i}(0) + q_{s}(l)}{\sqrt{2}} \right\rangle^{2} %= \frac{\langle {\mathrm{\Delta}q^2_i(0)} \rangle + \langle {\mathrm{\Delta}q^2_s(l)} \rangle + 2\left\lbrack {\langle {q_{i}(0)q_{s}(l)}\rangle - \langle {q_{i}(0)} \rangle\langle {q_{s}(l)} \rangle} \right\rbrack}{2} \nonumber\\
= \frac{\left\langle {\Delta q^2_i(0)} \right\rangle + \left\langle {\Delta q^2_s(l)} \right\rangle}{2},\label{eq:d1var}\\
\langle\Delta d^2_2\rangle&= \left\langle \left( \frac{p_{i}(0) + p_{s}(l)}{\sqrt{2}} \right)^{2} \right\rangle - \left\langle \frac{p_{i}(0) + p_{s}(l)}{\sqrt{2}} \right\rangle^{2} %= \frac{\left\langle {\mathrm{\Delta}p^2_i(0)} \right\rangle + \left\langle {\mathrm{\Delta}p^2_s(l)} \right\rangle + 2\left\lbrack {\left\langle {p_{i}(0)p_{s}(l)} \right\rangle - \left\langle {p_{i}(0)} \right\rangle\left\langle {p_{s}(l)} \right\rangle} \right\rbrack}{2} \nonumber\\
= \frac{\left\langle {\Delta p^2_i(0)} \right\rangle + \left\langle {\Delta p^2_s(l)} \right\rangle}{2}.
\label{eq:d2var}
\end{align}
\end{subequations}

Mathematically, Eqs.~(\ref{eq:d1var}) and (\ref{eq:d2var}) imply that the (logarithmic) noise characteristics of $\langle\Delta d^2_1\rangle$ and $\langle\Delta d^2_2\rangle$ can be inferred from the variances of single-mode quadratures, aided by Figs.~2(a)-(d) in the main text. Physically, we anticipate that both dynamical and stationary C2Q transitions can be also observed in $\langle\Delta d^2_1\rangle$, depending on the length $l$ of the medium.

\begin{figure}[htbp]
\centering
\fbox{\includegraphics[width=0.92\linewidth]{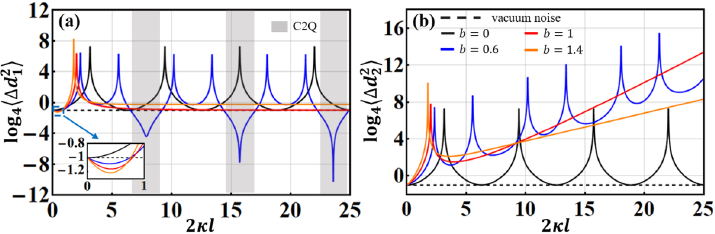}}
\caption{In comparisons to usual TMSV ($b=0$, black solid) and vacuum noise (black dashed), $d_1$ (a) and $d_2$ (b) exhibit distinct squeezing and de-squeezing features under type-II quadrature PT symmetry. In the PT phase-unbroken region ($b<1$), $\langle\Delta d^2_1\rangle$ displays dynamic C2Q transitions for some flexible lengths whereas stationary C2Q transitions at any specifically fixed length within the ranges (illustrated by gray shaded rectangular), demarcated by EP-variance curve as exact boundary between classical and quantum noise realms for the latter scenario. Conversely, $d_2$ consistently experiences classical noise amplification. For comparison, the dashed lines denote vacuum noise, while the solid black lines depict the standard two-mode squeezed vacuum without PSA and loss. The inset in (a) is a zoom-in view for the dimensionless propagation length $2\kappa l\in(0,1)$. Parameters remain consistent with those in Fig.~2 of the main text.}
\label{fig:Fig3}
\end{figure}

In comparison to Fig.~2 presented in the main text, the noise characteristics of the two-mode quadrature observables $d_1$ and $d_2$ (illustrated in Fig.~\ref{fig:Fig3}(a) and (b)) inherit, to some extent, those of the single-mode quadratures $\{q_i(0),p_s(l),q_s(l),p_i(0)\}$ but exhibit discernible differences. Akin to $\{q_i(0),q_s(l)\}$, dual transitions manifest in $d_1$: besides the regular $\pm\beta$-phase transition (Fig.~1(b) in the main text), a stationary C2Q transition takes place for some properly fixed lengths $l$ by changing $b$, with its EP-variance (red solid) curve acting as the partition boundary between classical and quantum noise distributions. For your reference, we have illustrated the emergence of these stationary C2Q transitions with gray shaded rectangles in Fig.~\ref{fig:Fig3}(a). (Please note that the stationary C2Q transitions manifested in single-mode quadratures $q_i(0)$ and $q_s(l)$ (Figs.~2(a) and (c) in the main text, as well as NF (Fig.~S2(a) below), can be similarly understood in this way.) Simultaneously, a dynamical C2Q transition emerges in the quadrature-PT phase intact region (blue solid) for a variable length, with the boundary curve dynamically responding to the $b$ value. Moreover, a slight noise reduction (zoom-in inset) is consistently visible over a very short propagation distance, regardless of PT symmetry, owing to the insufficient competition between PT and squeezing. Beyond this range, the logarithmic variance of $d_1$ proceeds with a more symmetrical shape compared to Figs.~2(a) and 2(c) in the main text and comprises interleaved dual periodic oscillations with maximum squeezing outputs at troughs ($l=n\pi/\beta$ with $n$ being positive integers) before PT symmetry spontaneously breaks down. However, it diminishes to a sole peak at $l=\epsilon/\beta$ with classical noise fluctuations on top of the vacuum-noise level (black dashed line) at and above the EP ($b=1$). On the other hand, as illustrated in Fig.~\ref{fig:Fig3}(b), $d_2$ always encounters the classical noisy amplification but showcases dissimilar noise undulation over different PT zones. Below the EP threshold (blue curve), $\log_4\langle\Delta d^2_2\rangle$ incrementally displays double periodic fluctuations at the same period $\mathrm{T}=2\pi\kappa/\beta$, with repeated double peaks occurring sequentially at $l=[(n-1)\pi+\epsilon]/\beta$ and $l'=(n\pi-\epsilon)/\beta$. At and above the EP, it ceases to a single peak centered at $l=\epsilon/\beta$, with its right-side curve flattening out as $b$ is further augmented. For reference, the variances of $d_1$ and $d_2$ for the standard TMSV ($b=0$, solid black curves) are also provided in Fig.~\ref{fig:Fig3}(a) and (b). As explained in our previous study on type-I quadrature PT symmetry \cite{1}, the decision to plotting logarithmic variances in this context is purely for the numerical convenience.

Before proceeding the discussion, let us make further remarks: Upon comparing our current findings with the type-I case (Figs. 3(a) and (b) in Ref. \cite{1}), significant deviations become apparent. The key disparity lies in the divergent C2Q transitions observed in $\log_4\langle\Delta d^2_1\rangle$. This discrepancy arises from the fact that in type-II (PSA-only), $d_1$ is ruled by both active and passive PT symmetry, whereas in type-I (balanced PSA and loss), $d_1$ is governed by active PT and the TMSV, thus leading to distinct noise profiles. This disparity also manifests the different curves of $\log_4\langle\Delta d^2_2\rangle$ under the two situations. These distinctions underscore the importance of considering the specific physical system and parameter selection space when analyzing the response of PT symmetry to quadrature squeezing.

\subsection{Relative intensity squeezing measurement}
In addition to homodyning detection, the relative intensity squeezing measurement (RISM) \cite{NF} provides an alternative method to explore quadrature PT symmetry and the C2Q transitions. As previously demonstrated \cite{1}, RISM offers an ingenious alternative for capturing the essential aspects of type-I quadrature PT symmetry, even in scenarios with two highly unbalanced channels. This detection technique cleverly allows the shot noise of one beam to be measured and subtracted from the other, resulting in a lower-noise differential measurement of the desired signal. In practice, this is often achieved by splitting the signal beam using a 50:50 beam splitter and then subtracting the photo-currents of the two outgoing beams using a balanced photo-detector. By properly adjusting the relative phase between the two outgoing beams, the noise level of the difference photo-current can be reduced below the shot noise limit, thence enabling squeezing in the relative intensity noise of the two beams. From this point of view, the RISM technique furnishes a complementary approach to standard quadrature homodyne detection for studying quadrature PT symmetry and the C2Q transitions in two-mode squeezed states.

Now we shift our attention to the relative-intensity operator, defined as $N_i(0)-N_s(l)=a^{\dagger}_i(0)a_i(0)-a^{\dagger}_s(l)a_s(l)$. This photon-number difference operator acts as a metric for quantifying the extent of squeezing in a PDC or FWM process. It capitalizes on the strong quantum correlation present in the synchronized photon number fluctuations in twin beams, which induce relative-intensity squeezing in the paired signal-idler modes. Derived from the $N_i(0)-N_s(l)$ operator, the noise figure (NF) emerges as a measurable indicator that characterizes the level of squeezing. NF represents the ratio of the measured relative-intensity noise to the combined shot noise of the two unequal fields. 

Mathematically, NF \cite{NF} is defined as:
\begin{eqnarray}
\mathrm{NF}\equiv\frac{\mathrm{Var}[N_i(0)-N_s(l)]}{\langle N_i(0)\rangle+\langle N_s(l)\rangle}=\frac{\langle\Delta N^2_i(0)\rangle+\langle\Delta N^2_s(l)\rangle-2[\langle N_i(0)N_s(l)\rangle-\langle N_i(0)\rangle\langle N_s(l)\rangle]}{\langle N_i(0)\rangle+\langle N_s(l)\rangle}.\label{eq:NF}
\end{eqnarray}
To evaluate the NF (\ref{eq:NF}), determining $a_{i}(0)$ and $a_{s}(l)$ in relation to their initial conditions are essential. Utilizing the key results described by Eqs. (2a) and (2b) in the main text, and following some algebraic manipulation, we arrive at their expressions as
\begin{subequations}
\begin{align}
a_{i}(0) = &e^{- \frac{gl}{2}}q_{i}(l)\csc(\epsilon + \beta l)\sin\epsilon - p_s(0)\csc(\epsilon + \beta l)\sin(\beta l)\nonumber\\
&+ ie^{\frac{gl}{2}}p_{i}(l)\csc(\epsilon - \beta l)\sin\epsilon - iq_{s}(0)\csc(\epsilon - \beta l)\sin(\beta l) \nonumber\\
= &Aq_{i}(l) + iBp_{i}(l) + iCq_{s}(0) + Dp_{s}(0) \nonumber\\
=& \frac{A + B}{2}\left\lbrack {q_{i}(l) + ip_{i}(l)} \right\rbrack + \frac{A - B}{2}\left\lbrack {q_{i}(l) - ip_{i}(l)} \right\rbrack 
+ i\frac{C - D}{2}\left\lbrack {q_{s}(0) + ip_{s}(0)} \right\rbrack + i\frac{C + D}{2}\left\lbrack {q_{s}(0) - ip_{s}(0)} \right\rbrack \nonumber\\
 =& \frac{A + B}{2}a_{i}(l) + \frac{A - B}{2}a_{i}^{\dagger}(l) + i\frac{C - D}{2}a_{s}(0) + i\frac{C + D}{2}a_{s}^{\dagger}(0), \label{eq:ai} \\
a_{s}(l) =& e^{-\frac{gl}{2}}q_{s}(0)\csc(\epsilon - \beta l)\sin\epsilon - p_{i}(l)\csc(\epsilon - \beta l)\sin(\beta l)\nonumber\\
&+ ie^{\frac{gl}{2}}p_{s}(0)\csc(\epsilon + \beta l)\sin\epsilon - iq_{i}(l)\csc(\epsilon + \beta l)\sin(\beta l) \nonumber\\&= iDq_{i}(l) + Cp_{i}(l) + Eq_{s}(0) + iFp_{s}(0) \nonumber\\
=& i\frac{D - C}{2}\left\lbrack {q_{i}(l) + ip_{i}(l)} \right\rbrack + i\frac{C + D}{2}\left\lbrack {q_{i}(l) - ip_{i}(l)} \right\rbrack + \frac{E + F}{2}\left\lbrack {q_{s}(0) + ip_{s}(0)} \right\rbrack + \frac{E - F}{2}\left\lbrack {q_{s}(0) - ip_{s}(0)} \right\rbrack \nonumber\\
=& i\frac{D - C}{2}a_{i}(l) + i\frac{C + D}{2}a_{i}^{\dagger}(l) + \frac{E + F}{2}a_{s}(0) + \frac{E - F}{2}a_{s}^{\dagger}(0), \label{eq:as}
\end{align}
\end{subequations}
for the vacuum input. Here, the involved coefficients correspond to 
\begin{eqnarray}
&A=e^{-\frac{gl}{2}}\csc(\epsilon + \beta l)\sin\epsilon,\; B=e^{\frac{gl}{2}}\csc(\epsilon - \beta l)\sin\epsilon,\nonumber\\
&C=- \csc(\epsilon - \beta l)\sin(\beta l), \; D=- \csc(\epsilon + \beta l)\sin(\beta l),\nonumber\\
&E=e^{-\frac{gl}{2}}\csc(\epsilon - \beta l)\sin\epsilon, F=e^{\frac{gl}{2}}\csc(\epsilon + \beta l)\sin\epsilon.\nonumber
\end{eqnarray} 

Using Eqs.~(\ref{eq:ai}) and (\ref{eq:as}), we next compute the terms involved in the definition of NF (\ref{eq:NF}): $\left\langle\Delta N_{i}^{2}(0)\right\rangle$, $\left\langle\Delta N_{s}^{2}(l)\right\rangle$, and $\left\langle {N_{i}(0)N_{s}(l)}\right\rangle - \left\langle {N_{i}(0)} \right\rangle\left\langle {N_{s}(l)} \right\rangle$. After some calculations, we have
\begin{subequations}
\begin{align}
\left\langle {\Delta N_{i}^{2}(0)} \right\rangle &= \left\langle {a_{i}^{\dagger}(0)a_{i}(0)a_{i}^{\dagger}(0)a_{i}(0)} \right\rangle - \left\langle {a_{i}^{\dagger}(0)a_{i}(0)} \right\rangle^{2} \nonumber\\&= \frac{1}{8}\left\lbrack {\left( {A^{2} + D^{2}} \right)^{2} + \left( {B^{2} + C^{2}} \right)^{2} - 2\left( {AB - CD} \right)^{2}} \right\rbrack \nonumber\\&= \frac{1}{8}\left\lbrack {\left( {A^{2} + D^{2}} \right)^{2} + \left( {B^{2} + C^{2}} \right)^{2} - 2} \right\rbrack, \label{eq:Nivar} \\
\left\langle {\Delta N_{s}^{2}(l)} \right\rangle &= \left\langle {a_{s}^{\dagger}(l)a_{s}(l)a_{s}^{\dagger}(l)a_{s}(l)} \right\rangle - \left\langle {a_{s}^{\dagger}(l)a_{s}(l)} \right\rangle^{2} \nonumber\\&= \frac{1}{8}\left\lbrack {\left( {F^{2} + D^{2}} \right)^{2} + \left( {E^{2} + C^{2}} \right)^{2} - 2\left( {FE - CD} \right)^{2}} \right\rbrack \nonumber\\&= \frac{1}{8}\left\lbrack {\left( {F^{2} + D^{2}} \right)^{2} + \left( {E^{2} + C^{2}} \right)^{2} - 2} \right\rbrack, \label{eq:Nsvar} \\
\left\langle {N_{i}(0)N_{s}(l)} \right\rangle - \left\langle {N_{i}(0)} \right\rangle\left\langle {N_{s}(l)} \right\rangle& = \left\langle {a_{i}^{\dagger}(0)a_{i}(0)a_{s}^{\dagger}(l)a_{s}(l)} \right\rangle - \left\langle {a_{i}^{\dagger}(0)a_{i}(0)} \right\rangle\left\langle {a_{s}^{\dagger}(l)a_{s}(l)} \right\rangle \nonumber\\&= \frac{1}{8}\left\lbrack {\left( {A^{2} - B^{2} - E^{2} + F^{2}} \right)\left( {D^{2} - C^{2}} \right) + 2\left( {AD + BC} \right)\left( {CE + DF} \right)} \right\rbrack \nonumber\\&= \frac{1}{8}\left\lbrack {\left( {AD + DF} \right)^{2} + \left( {CE + BC} \right)^{2} - \left( {AC - DE} \right)^{2} - \left( {BD - FC} \right)^{2}} \right\rbrack \nonumber\\&= \frac{1}{8}\left\lbrack {\left( {AD + DF} \right)^{2} + \left( {CE + BC} \right)^{2}} \right\rbrack. \label{eq:NiNscovar}
\end{align}
\end{subequations}
The commutation relations (\ref{eq:commutation relations_qipi}) and (\ref{eq:commutation relations_qsps}) below imply the following useful identities: $ AB-CD = FE-CD = 1 $  and $ AC-DE = BD-FC = 0 $. Leveraging these results, after lengthy derivations, we arrive at the final NF formula below, albeit complicated,
\begin{align}
\mathrm{NF}= \frac{\left\lbrack {\left( {A^{2} + D^{2}} \right)^{2} + \left( {B^{2} + C^{2}} \right)^{2} + \left( {F^{2} + D^{2}} \right)^{2} + \left( {E^{2} + C^{2}} \right)^{2} - 4 - 2D^{2}\left( {A + F} \right)^{2} - 2C^{2}\left( {E + B} \right)^{2}} \right\rbrack}{2\left\lbrack {\left( {A - B} \right)^{2} + \left( {E - F} \right)^{2} + 2\left( {C + D} \right)^{2}} \right\rbrack}. \label{eq:NF_detail}
\end{align}
Note that the $A$--$B$ coefficients have been given right after Eqs.~(\ref{eq:ai}) and (\ref{eq:as}).

\begin{figure}[htbp]
\centering
\fbox{\includegraphics[width=0.95\linewidth]{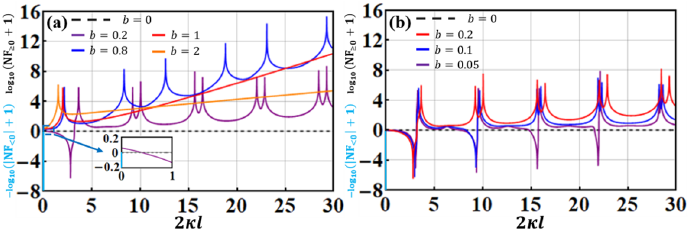}}
\caption{PT-modulated NF in relative intensity squeezing measurement. In (a), the logarithmic NF illustrates classical noise amplification for $b>0.61$, whereas relative-intensity squeezing is observed for $0<b<0.6$ at short interaction lengths. Moreover, for $b\geqslant1$, the logarithmic NF exhibits dual periodic oscillations with sequential peaks at $2\kappa(n\pi-\epsilon)/\beta$ and $2\kappa[(n-1)\pi+\epsilon]/\beta$ ($n\in\mathbb{N}^{+}$, i.e., positive integers). However, when $b<1$, the logarithmic NF manifests as a solitary peak above the vacuum noise level. As exemplified in (b), quantum squeezing occurs when $b\leqslant0.6$, persisting over longer distances for smaller $b$ values before disappearing. The dashed lines represent the regular two-mode squeezed vacuum without PSA and loss. The inset in (a) is a zoom-in view for the dimensionless propagation length $2\kappa l\in(0,1)$.}
\label{fig:NF}
\end{figure}

Since the RISM makes use of the relative-intensity fluctuation rather than simpler quadrature variances as the quantum observable, we anticipate peculiar noise statistics may arise from the NF. Our numerical simulations indeed confirm this expectation. From Figs.~\ref{fig:NF}(a) and (b), we observe distinct NF curve patterns before and after the regular PT-phase transition. Specifically, in the PT-phase intact regime ($b<1$), the logarithmic NF gives rise to dual periodic oscillations with the same periodicity $T=2\pi\kappa/\beta$ and features two adjacent peaks at $2\kappa(n\pi-\epsilon)/\beta$ and $2\kappa[(n-1)\pi+\epsilon]/\beta$. In this region, negative values signify quantum squeezing occurrences. Through numerical simulations, we find that NF no longer strictly remains below 0 when $b$ exceeds $0.6\sim0.61$, due to channel imbalances. Thus, Fig.~\ref{fig:NF} suggests that for $b\in(0,0.6)$, quantum squeezing can persist for some evolution distance before transitioning into classical noise distributions. 

When $b\geqslant1$, $\lg(\mathrm{NF}+1)$ demonstrates steady classical noise amplification with a solitary peak at $2\ln(\sqrt{b^2-1}+b)/{\sqrt{b^2-1}}$, as seen in Fig.~\ref{fig:NF}(a). This contrasts with our recent findings in type-I quadrature PT symmetry \cite{1}, where a balanced loss rate $\gamma$ introduced into the signal path disrupted the system's symmetry, bringing all values of $\lg(\mathrm{NF}+1)$ into the upper space when $\gamma/\kappa\geqslant0.53$. This alternative method suggests another approach to identify regular PT-phase transitions, in addition to the typical means of extracting eigenvalues. By observing the sharp change in NF patterns while manipulating $b$ for a fixed appropriate length, one can deduce the occurrence of the PT phase transition indirectly. Note that this indirect method is also applicable to single-mode and two-mode quadrature measurements. 

Moreover, both dynamical and stationary C2Q transitions are also achievable in RISM. For a small fixed length, the stationary C2Q transition in relative-intensity noise fluctuations can be realized by simply varying $b$, with the upper boundary at $b=0.6\sim0.61$. Conversely, for a slightly larger length, the dynamical C2Q transition can be approached with the separation set by $b=0.6\sim0.61$. In short, the RISM approach offers a neat and elegant solution to explore type-II quadrature-PT symmetric systems without classical analogies. 

To accurately delineate the classical and quantum noise domains, we opt to numerically plot the NF using the function $\lg(\mathrm{NF}_{\geqslant0}+1)$ for $\mathrm{NF}\geqslant0$ and $-\lg(\vert\mathrm{NF}_{<0}\vert+1)$ for $\mathrm{NF}<0$, as these logarithm expressions effectively represent the original noise contours and allow us to depict the full patterns more conveniently. The same strategy has been previously applied to the study of type-I quadrature PT symmetry \cite{1}. 

\subsection{Verification of the commutation relations}
In line with our previous research \cite{1}, it is important to validate the commutation relationships of the solutions (2a) and (2b) presented in the main text to ensure the consistency of our findings. Demonstrating their adherence to the requisite commutation relations is a straightforward task. 

By working out the algebra, one can show that the solutions (2a) and (2b) in the main text indeed satisfy the demanded commutation relations:
\begin{subequations}
\begin{align}
\left\lbrack {q_{i}(0),p_{i}(0)} \right\rbrack &= \frac{\sin^{2}\epsilon}{{\sin(\epsilon + \beta l)}{\sin(\epsilon - \beta l)}}\left\lbrack {q_{i}(l),p_{i}(l)} \right\rbrack + \frac{\sin^{2}(\beta l)}{{\sin(\epsilon + \beta l)}{\sin(\epsilon - \beta l)}}\left\lbrack {p_{s}(0),q_{s}(0)} \right\rbrack = \frac{i}{2}, \label{eq:commutation relations_qipi} \\
\left\lbrack {q_{s}(l),p_{s}(l)} \right\rbrack &= \frac{\sin^{2}(\beta l)}{{\sin(\epsilon + \beta l)}{\sin(\epsilon - \beta l)}}\left\lbrack {p_{i}(l),q_{i}(l)} \right\rbrack + \frac{\sin^{2}\epsilon}{{\sin(\epsilon + \beta l)}{\sin(\epsilon - \beta l)}}\left\lbrack {q_{s}(0),p_{s}(0)} \right\rbrack = \frac{i}{2}. \label{eq:commutation relations_qsps}
\end{align}
\end{subequations}
Alternatively, confirming the preservation of the correct commutation relations (\ref{eq:commutation relations_qipi}) and (\ref{eq:commutation relations_qsps}) substantiates the accuracy of Eqs.~(1a) and (1b) given in the main text.

\subsection{Type-II quadrature PT symmetry at phase angle $\varphi = \frac{\pi}{2}$}
So far, our discussions have been carried out under the assumption of a zero squeezing angle within the Hamiltonian $H = i\hslash g(a_{i}^{2} - a_{i}^{\dagger 2})/2 + \hslash\kappa(a_{i}^{\dagger}a_{s}^{\dagger} + a_{i}a_{s})$ as given in the main text. Typically, this Hamiltonian is associated with a variable squeezing angle $\varphi$,
\begin{align}
H_{\varphi} = i\hslash g(a_{i}^{2} - a_{i}^{\dagger 2})/2 + \hslash\kappa(a_{i}^{\dagger}a_{s}^{\dagger}e^{-i\varphi} + a_{i}a_{s}e^{i\varphi}), \label{eq:H_detail}
\end{align}
originating from the input pump. Here, $\varphi$ represents the overall phase of the input pump fields engaged in the nonlinear FWM process. 

In the following, we would like to examine an alternative case where $\varphi = \frac{\pi}{2}$ rather than 0 and see its impact on the phenomena of the proposed quadratrure PT symmetry. By following a similar procedure, we can derive two sets of coupled-quadrature equations:
\begin{subequations}
\begin{align}
\frac{d}{dz}\begin{bmatrix}
q_{i} \\
q_{s}
\end{bmatrix} &= \begin{bmatrix}
g & \kappa \\
{- \kappa} & 0
\end{bmatrix}\begin{bmatrix}
q_{i} \\
q_{s}
\end{bmatrix}, \label{eq:equation_qiqs} \\
\frac{d}{dz}\begin{bmatrix}
p_{i} \\
p_{s}
\end{bmatrix} &= \begin{bmatrix}
{- g} & {- \kappa} \\
\kappa & 0
\end{bmatrix}\begin{bmatrix}
p_{i} \\
p_{s}
\end{bmatrix}. \label{eq:equation_pips}
\end{align}
\end{subequations}
From Eqs.(\ref{eq:equation_qiqs}) and (\ref{eq:equation_pips}), by the similar procedure we can obtain the effective NH Hamiltonian matrices that govern the dynamics of the two sets of quadrature pairs $\left\{ {q_{i},q_{s}} \right\}$  and $\left\{ {p_{i},p_{s}} \right\}$:
\begin{subequations}
\begin{align}
H_{({q_{i},q_{s}})} &= \begin{bmatrix}
{ig} & {i\kappa} \\
{- i\kappa} & 0
\end{bmatrix} = i\frac{g}{2}\mathbb{I} + \begin{bmatrix}
{i\frac{g}{2}} & {i\kappa} \\
{- i\kappa} & {- i\frac{g}{2}}
\end{bmatrix} = i\frac{g}{2}\mathbb{I} + H'_{a}, \label{eq:H_qiqs} \\
H_{({p_{i},p_{s}})} &= \begin{bmatrix}
{- ig} & {i\kappa} \\
{- i\kappa} & 0
\end{bmatrix} = - i\frac{g}{2}\mathbb{I} + \begin{bmatrix}
{- i\frac{g}{2}} & {- i\kappa} \\
{i\kappa} & {i\frac{g}{2}}
\end{bmatrix} = - i\frac{g}{2}\mathbb{I} + H'_{p}. \label{eq:H_pips}
\end{align}
\end{subequations}
In contrast to the case where $\varphi=0$ (see Eqs.~(\ref{eq:H_qips}) and (\ref{eq:H_piqs}) in Section~1.A), we observe that here $H'_{a} + H'_{p} = 0$. Additionally, both quadrature pairs evolve PT-symmetrically, satisfying $\left\lbrack {H_{({q_{i},q_{s}})},PT} \right\rbrack = \left\lbrack {H_{({p_{i},p_{s}})},PT} \right\rbrack = 0$. Similar to the $\varphi=0$ case, now $\left\{ {q_{i},q_{s}} \right\}$ follows active PT symmetry, while
$\left\{ {p_{i},p_{s}} \right\}$ exhibits passive PT symmetry. When comparing the two cases -- $\varphi=0$ and $\varphi=\frac{\pi}{2}$, it is interesting to note that, owing to the symmetry inherent in the type-II system, the initial overall pump phase $\varphi$ rotates the quadrature pairs and provides additional options for establishing dual quadrature PT symmetry. This flexibility is unique to our scheme and may become challenging for other PT configurations. 

It is also worthwhile to point out that the original Hamiltonian~(\ref{eq:H_detail}) of the system is in fact Hermitian with respect to $a_i$ and $a_s$. The non-Hermiticity appearing in the resulting effective Hamiltonians (\ref{eq:H_qiqs}) and (\ref{eq:H_pips}) as well as the effective Hamiltonians (\ref{eq:H_qips}) and (\ref{eq:H_piqs}) stems from the evolution of the corresponding quadratures. This is also applicable to type-I quadrature PT symmetry \cite{1}. One indication of non-Hermiticity in these derived effective Hamiltonian matrices is the appearance of the EP(s), which strongly signifies anti-Hermiticity. EPs differ from the degeneracies observed in Hermitian Hamiltonians. The emergence of EPs resembles scenarios of perfect anti-PT symmetry, even in the absence of gain and loss components \cite{3,4,5,6,7,8,at9,at10,at11,at12}.

With use of the initial boundary conditions, the general solutions to Eqs.~(\ref{eq:equation_qiqs}) and (\ref{eq:equation_pips}) take on the following compact and symmetrical forms:
\begin{subequations}
\begin{align}
\begin{bmatrix}
{q_{i}(0)} \\
{q_{s}(l)}
\end{bmatrix} &= {\csc\left( {\beta l + \epsilon} \right)}\begin{bmatrix}
{e^{\frac{- gl}{2}}{\sin\epsilon}} & {- {\sin(\beta l)}} \\
{- {\sin(\beta l)}} & {e^{\frac{gl}{2}}{\sin\epsilon}}
\end{bmatrix}\begin{bmatrix}
{q_{i}(l)} \\
{q_{s}(0)}
\end{bmatrix}, \label{eq:qiqs_II} \\
\begin{bmatrix}
{p_{i}(0)} \\
{p_{s}(l)}
\end{bmatrix} &= {\csc\left( {\beta l - \epsilon} \right)}\begin{bmatrix}
{- e^{\frac{gl}{2}}{\sin\epsilon}} & {- {\sin(\beta l)}} \\
{- {\sin(\beta l)}} & {- e^{\frac{- gl}{2}}{\sin\epsilon}}
\end{bmatrix}\begin{bmatrix}
{p_{i}(l)} \\
{p_{s}(0)}
\end{bmatrix}. \label{eq:pips_II}
\end{align}
\end{subequations}
Again, it is not difficult to check that the solutions (\ref{eq:qiqs_II}) and (\ref{eq:pips_II}) satisfy the necessary commutation relations:
\begin{subequations}
\begin{align}
\left\lbrack {q_{i}(0),p_{i}(0)} \right\rbrack &= \frac{- {\sin^{2}\epsilon}}{{\sin\left( {\beta l + \epsilon} \right)}{\sin\left( {\beta l - \epsilon} \right)}}\left\lbrack {q_{i}(l),p_{i}(l)} \right\rbrack + \frac{\sin^{2}{\beta l}}{{\sin\left( {\beta l + \epsilon} \right)}{\sin\left( {\beta l - \epsilon} \right)}}\left\lbrack {q_{s}(0),p_{s}(0)} \right\rbrack = \frac{i}{2}, \label{eq:commutation relations_qipi_II} \\
\left\lbrack {q_{s}(l),p_{s}(l)} \right\rbrack &= \frac{\sin^{2}{\beta l}}{{\sin\left( {\beta l + \epsilon} \right)}{\sin\left( {\beta l - \epsilon} \right)}}\left\lbrack {q_{i}(l),p_{i}(l)} \right\rbrack + \frac{- {\sin^{2}\epsilon}}{{\sin\left( {\beta l + \epsilon} \right)}{\sin\left( {\beta l - \epsilon} \right)}}\left\lbrack {q_{s}(0),p_{s}(0)} \right\rbrack = \frac{i}{2}. \label{eq:commutation relations_qsps_II}
\end{align}
\end{subequations}

As these results are essentially no fundamental difference from those obtained in the $\varphi=0$ case, all our discussions pertaining to the latter case are applicable to the $\varphi=\frac{\pi}{2}$ scenario. Regarding other angles $\varphi$, we leave their exploration as an exercise for the reader.

\section{EPR correlations and entanglement in type-II quadrature PT symmetry}
In what follows, we would like to make an effort on investigating the nontrivial connection between nonclassical correlations (including quantum entanglement) and PT symmetry utilizing our type-II quadrature PT system. We are aware that the area of research remains largely unexplored, especially in the continuous-variable (CV) domain \cite{CV1,CV2,CV3,CV4}. It is worth to point out that the discussion presented in this section provides additional insights into quadrature PT symmetry, in addition to the nontrivial C2Q transitions studied above.

\subsection{Calculations of quantum-mechanical correlation coefficients}
Historically, measurements of quadrature phase and amplitude of two-mode squeezed light have been widely employed in the study of CV entanglement \cite{9,10,su,duan,simon}. To rigorously test entanglement, the two modes may be separately or jointly detected, and correlation measurements be performed on the photo-currents. One of the methods for characterizing nonclassicality and quantum entanglement, proposed by Reid \cite{9,10} using quantum-mechanical correlation coefficients, follows the EPR standard. 

Thanks to the simplicity and idealism of the type-II configuration, it provides us a quintessential platform for exploring the peculiar relationship between nonclassical correlation and PT symmetry from a theoretical point of view. Utilizing Eqs.~(2a)-(2b) and (4) given in the main text, after some labor, we get the following interesting expressions for the quantum-mechanical correlation coefficients $\left| C_{j,m} \right|$ with $\left\{ {j,m} \right\}=\left\{ {1,2} \right\}$ for the case where $\varphi=0$:
\begin{equation}
\left| C^{\varphi=0}_{11} \right|= \left| \frac{\left\langle {X_{1}(0)Y_{1}(l)} \right\rangle}{\sqrt{\left\langle {X^{2}_{1}(0)} \right\rangle\left\langle {Y^{2}_{1}(l)}\right\rangle}} \right| = \left| \frac{\left\langle {q_{i}(0)q_{s}(l)} \right\rangle}{\sqrt{\left\langle {q^{2}_{i}(0)} \right\rangle\left\langle {q^{2}_{s}(l)} \right\rangle}} \right| = 0, \label{eq:correlation coefficients_11_0} 
\end{equation}
\begin{equation}
\left| C^{\varphi=0}_{22} \right| = \left| \frac{\left\langle {X_{2}(0)Y_{2}(l)} \right\rangle}{\sqrt{\left\langle {X^{2}_{2}(0)} \right\rangle\left\langle {Y^{2}_{2}(l)} \right\rangle}} \right| = \left| \frac{\left\langle {p_{i}(0)p_{s}(l)} \right\rangle}{\sqrt{\left\langle {p^{2}_{i}(0)} \right\rangle\left\langle {p^{2}_{s}(l)} \right\rangle}} \right| = 0, \label{eq:correlation coefficients_22_0}
\end{equation}
\begin{align}
\left| C^{\varphi=0}_{12} \right| &= \left| \frac{\left\langle {X_{1}(0)Y_{2}(l)} \right\rangle}{\sqrt{\left\langle {X^{2}_{1}(0)} \right\rangle\left\langle {Y^{2}_{2}(l)} \right\rangle}} \right| = \left| \frac{\left\langle {q_{i}(0)p_{s}(l)} \right\rangle}{\sqrt{\left\langle {q^{2}_{i}(0)} \right\rangle\left\langle {p^{2}_{s}(l)} \right\rangle}} \right|\nonumber\\
&= \left|\frac{-e^{-\frac{gl}{2}}\csc^{2}(\beta l+\epsilon)\sin\epsilon\sin(\beta l)\left\langle q^{2}_{i}(l) \right\rangle - e^{\frac{gl}{2}}\csc^{2}(\beta l+\epsilon)\sin\epsilon\sin(\beta l)\left\langle p^{2}_{s}(0) \right\rangle}{\sqrt{\left[ \frac{e^{-gl}{\sin^2\epsilon} + {\sin^2(\beta l)}}{4{\sin}^{2}(\beta l + \epsilon)} \right]\left[\frac{e^{gl}{\sin^2\epsilon} + \sin^2(\beta l)}{4\sin^2(\beta l + \epsilon)}\right]}} \right| \nonumber\\
&= \left| \frac{e^{- \frac{gl}{2}}{\mathit{\sin}\epsilon}{\mathit{\sin}(\beta l)} + e^{\frac{gl}{2}}{\mathit{\sin}\epsilon}{\mathit{\sin}(\beta l)}}{\sqrt{\left[ {e^{- gl}{\mathit{\sin}^{2}\epsilon} + {\mathit{\sin}^{2}(\beta l)}} \right]\left[ {e^{gl}{\mathit{\sin}^{2}\epsilon} + {\mathit{\sin}^{2}(\beta l)}} \right]}} \right| \nonumber\\
&= 2{\cosh\left(\frac{gl}{2}\right)}\left| \frac{{\mathit{\sin}\epsilon}{\mathit{\sin}(\beta l)}}{\sqrt{{\mathit{\sin}^{4}\epsilon} + {\mathit{\sin}^{4}(\beta l)} + 2\cosh(gl){\mathit{\sin}^{2}\epsilon}{\mathit{\sin}^{2}(\beta l)}}} \right|, \label{eq:correlation coefficients_12_0}
\end{align}
and
\begin{align}
\left| C^{\varphi=0}_{21} \right| &= \left| \frac{\left\langle {X_{2}(0)Y_{1}(l)} \right\rangle}{\sqrt{\left\langle {X^{2}_{2}(0)} \right\rangle\left\langle {Y^{2}_{1}(l)} \right\rangle}} \right| = \left| \frac{\left\langle {p_{i}(0)q_{s}(l)} \right\rangle}{\sqrt{\left\langle {p^{2}_{i}(0)} \right\rangle\left\langle {q^{2}_{s}(l)} \right\rangle}} \right| \nonumber\\
&= \left| \frac{- e^{\frac{gl}{2}}{\csc}^{2}\left( {\beta l - \epsilon} \right){\mathit{\sin}\epsilon}{\mathit{\sin}(\beta l)}\left\langle {p^{2}_{i}(l)} \right\rangle - e^{- \frac{gl}{2}}{\csc}^{2}\left( {\beta l - \epsilon} \right){\mathit{\sin}\epsilon}{\mathit{\sin}(\beta l)}\left\langle {q^{2}_{s}(0)} \right\rangle}{\sqrt{\left[ \frac{e^{- gl}{\mathit{\sin}^{2}\epsilon} + {\mathit{\sin}^{2}(\beta l)}}{4{\sin}^{2}\left({\beta l - \epsilon} \right)} \right]\left[ \frac{e^{gl}{\mathit{\sin}^{2}\epsilon} + {\mathit{\sin}^{2}(\beta l)}}{4{\sin}^{2}\left( {\beta l - \epsilon} \right)} \right]}} \right| \nonumber\\
&= \left| \frac{e^{- \frac{gl}{2}}{\mathit{\sin}\epsilon}{\mathit{\sin}(\beta l)} + e^{\frac{gl}{2}}{\mathit{\sin}\epsilon}{\mathit{\sin}(\beta l)}}{\sqrt{\left[ {e^{- gl}{\mathit{\sin}^{2}\epsilon} + {\mathit{\sin}^{2}(\beta l)}} \right]\left[ {e^{gl}{\mathit{\sin}^{2}\epsilon} + {\mathit{\sin}^{2}(\beta l)}} \right]}} \right| \nonumber\\
&= 2{\cosh\left(\frac{gl}{2}\right)}\left| \frac{{\mathit{\sin}\epsilon}{\mathit{\sin}(\beta l)}}{\sqrt{{\mathit{\sin}^{4}\epsilon} + {\mathit{\sin}^{4}(\beta l)} + 2\cosh(gl){\mathit{\sin}^{2}\epsilon}{\mathit{\sin}^{2}(\beta l)}}} \right|. \label{eq:correlation coefficients_21_0}
\end{align}

These results have been presented in the main text as Eqs.~(5a) and (5b). From the above findings, we find that due to the absence of coupling for the quadrature pairs $\left\{ {q_{i}(0),q_{s}(l)} \right\}$ and $\left\{ {p_{i}(0),p_{s}(l)} \right\}$, their quantum correlation coefficients (\ref{eq:correlation coefficients_11_0}) and (\ref{eq:correlation coefficients_22_0}) are always zero. However, this does not hold true for $\left\{ {q_{i}(0),p_{s}(l)} \right\}$ and $\left\{ {p_{i}(0),q_{s}(l)} \right\}$. On the other hand, since $H_{(q_{i},p_{s})}$ (\ref{eq:H_qips}) and $H_{(p_{i},q_{s})}$ (\ref{eq:H_piqs}) are complementary, i.e., $H_a + H_p^{\top} = 0$, their quantum correlation coefficients (\ref{eq:correlation coefficients_12_0}) and (\ref{eq:correlation coefficients_21_0}) are identical. This nonzero quantum correlation coefficient allows for an EPR reasoning, as emphasized in the main text. As explicitly illustrated in Fig.~3(a) in the main text, the nonclassicality behaves distinctively before and after the PT-phase transition at the EP ($b=1$). Interestingly, we observe that quantum correlation coefficients (\ref{eq:correlation coefficients_12_0}) and (\ref{eq:correlation coefficients_21_0}) provide distinct insights into understanding the nonclassicality of the system compared to the single-mode or two-mode quadrature variances, as well as the relative-intensity squeezing aforementioned.Apart from these observations, it might be interesting to explore the nontrivial connection with quantum steering \cite{CV4,steering1}, examining how PT symmetry manifests in steering, nonlocaility, and entanglement.

By conducting a similar analysis for the case where $\varphi = \frac{\pi}{2}$ and using Eqs.~(\ref{eq:qiqs_II}) and (\ref{eq:pips_II}), we find that the quantum correlation coefficients accordingly turn to be $\lvert C^{\varphi=\pi/2}_{11}\rvert=\lvert C^{\varphi=\pi/2}_{22}\rvert=\lvert C^{\varphi=0}_{12}\rvert=\lvert C^{\varphi=0}_{21}\rvert$ and $C^{\varphi=\pi/2}_{12}=C^{\varphi=\pi/2}_{21}=C^{\varphi=0}_{11}=C^{\varphi=0}_{22}=0$, exactly opposite to the $\varphi=0$ case.

\subsection{Calculations of the EPR correlations}
For quite some time, it has been understood that quantum entanglement demands much stronger nonclassical correlations. In the CV framework, a variety of methods \cite{CV1,CV2,CV3,CV4,9,10} has been put forward to characterize these quantum correlations and entanglement. Alongside the quantum-mechanical correlation coefficients discussed in Section~2.A, one can also employ both output beams to demonstrate the system's high nonclassical properties. 

As such, we are interested in using the sum and difference combinations of the amplitude and phase quadratures of idler and signal fields defined in the main text -- $X^{\theta}_1(0)+Y^{\phi}_1(l)$, $X^{\theta}_1(0)-Y^{\phi}_1(l)$, $X^{\theta+\frac{\pi}{2}}_2(0)-Y^{\phi+\frac{\pi}{2}}_2(l)$, and $X^{\theta+\frac{\pi}{2}}_2(0)+Y^{\phi+\frac{\pi}{2}}_2(l)$ -- to reveal the intriguing relationship between PT symmetry and EPR entanglement. Here, we introduce our notions:
\begin{subequations}
\begin{align}
X_{1}^{\theta}(0) &= \frac{1}{2}\left\lbrack {e^{i\theta}a_{i}^{\dagger}(0) + e^{- i\theta}a_{i}(0)} \right\rbrack = q_{i}(0){\cos\theta} + p_{i}(0){\sin\theta}, \label{eq:X_1_theta} \\
X_{2}^{\theta + \frac{\pi}{2}}(0) &= \frac{i}{2}\left\lbrack {e^{i\theta}a_{i}^{\dagger}(0) - e^{- i\theta}a_{i}(0)} \right\rbrack = - q_{i}(0){\sin\theta} + p_{i}(0){\cos\theta}, \label{eq:X_2_theta} \\
Y_{1}^{\phi}(l) &= \frac{1}{2}\left\lbrack {e^{i\phi}a_{s}^{\dagger}(l) + e^{- i\phi}a_{s}(l)} \right\rbrack = q_{s}(l){\cos\phi} + p_{s}(l){\sin\phi}, \label{eq:Y_1_theta} \\
Y_{2}^{\phi + \frac{\pi}{2}}(l) &= \frac{i}{2}\left\lbrack {e^{i\phi}a_{s}^{\dagger}(l) - e^{- i\phi}a_{s}(l)} \right\rbrack = - q_{s}(l){\mathit{\sin}\phi} + p_{s}(l){\mathit{\cos}\phi}. \label{eq:Y_2_theta}
\end{align}
\end{subequations}

One can easily check that the amplitude quadratures sum and phase quadratures difference (or the amplitude quadratures difference and phase quadaratures sum) are commutative since $\left[X^{\theta}_1(0)+Y^{\phi}_1(l),X^{\theta+\frac{\pi}{2}}_2(0)-Y^{\phi+\frac{\pi}{2}}_2(l)\right]=0$ (or $\left[X^{\theta}_1(0)-Y^{\phi}_1(l),X^{\theta+\frac{\pi}{2}}_2(0)+Y^{\phi+\frac{\pi}{2}}_2(l)\right]=0$); while the amplitude quadratures sum and phase quadratures sum (or the amplitude quadratures difference and phase quadratures difference) between the two beams form one pair of Heisenberg uncertainty conjugate variables, as they satisfy the commutation relation $\left[X^{\theta}_1(0)+Y^{\phi}_1(l),X^{\theta+\frac{\pi}{2}}_2(0)+Y^{\phi+\frac{\pi}{2}}_2(l)\right]=i$ (or $\left[X^{\theta}_1(0)-Y^{\phi}_1(l),X^{\theta+\frac{\pi}{2}}_2(0)-Y^{\phi+\frac{\pi}{2}}_2(l)\right]=i$). 

In order to unveil quantum entanglement, the two generalized quadratures $\left\{ X^{\theta}_1(0)+Y^{\phi}_1(l),X^{\theta+\frac{\pi}{2}}_2(0)-Y^{\phi+\frac{\pi}{2}}_2(l) \right\}$ or $\left\{ X^{\theta}_1(0)-Y^{\phi}_1(l),X^{\theta+\frac{\pi}{2}}_2(0)+Y^{\phi+\frac{\pi}{2}}_2(l) \right\}$ must fulfill the inseparability criterion. The stronger EPR criterion demands the variances of these generalized quadrature pairs to satisfy
\begin{subequations}
\begin{align}
\mathrm{ET_1}\equiv\mathrm{Var}\left\lbrack{X_{1}^{\theta}(0) - Y_{1}^{\phi}(l)}\right\rbrack + \mathrm{Var}\left\lbrack{X_{2}^{\theta + \frac{\pi}{2}}(0) + Y_{2}^{\phi + \frac{\pi}{2}}(l)}\right\rbrack < \frac{1}{2}, \label{eq:strong_1} \\
\mathrm{ET_2}\equiv\mathrm{Var}\left\lbrack{X_{1}^{\theta}(0) + Y_{1}^{\phi}(l)}\right\rbrack + \mathrm{Var}\left\lbrack{X_{2}^{\theta + \frac{\pi}{2}}(0) - Y_{2}^{\phi + \frac{\pi}{2}}(l)}\right\rbrack < \frac{1}{2}. \label{eq:strong_2} 
\end{align}
\end{subequations}
Accordingly, the weaker EPR criterion requires these generalized quadrature pairs to obey
\begin{subequations}
\begin{align}
\mathrm{ET_1}\equiv\mathrm{Var}\left\lbrack{X_{1}^{\theta}(0) - Y_{1}^{\phi}(l)}\right\rbrack + \mathrm{Var}\left\lbrack{X_{2}^{\theta + \frac{\pi}{2}}(0) + Y_{2}^{\phi + \frac{\pi}{2}}(l)}\right\rbrack < 1, \label{eq:weaker_1} \\
\mathrm{ET_2}\equiv\mathrm{Var}\left\lbrack{X_{1}^{\theta}(0) + Y_{1}^{\phi}(l)}\right\rbrack + \mathrm{Var}\left\lbrack{X_{2}^{\theta + \frac{\pi}{2}}(0) - Y_{2}^{\phi + \frac{\pi}{2}}(l)}\right\rbrack < 1. \label{eq:weaker_2} 
\end{align}
\end{subequations}
It is notable that the inseparability criterion proposed by Duan \textit{et al}. \cite{duan} and Simon \cite{simon} falls within the scope of the weaker EPR criterion outlined above.

As noted from the above inequalities, for simplicity, we have denoted the entire left-hand side of Eq.~(\ref{eq:strong_1}) or (\ref{eq:weaker_1}) as ``$\mathrm{ET_1}$" and the whole left-hand side of Eq.~(\ref{eq:strong_2}) or (\ref{eq:weaker_2}) as ``$\mathrm{ET_2}$". Our goal now is to evaluate these variances for the input vacuum state. After some tedious derivations, we attain
\begin{subequations}
\begin{align}
4U_1U_2\cdot\mathrm{Var}\left\lbrack{X_{1}^{\theta}(0) - Y_{1}^{\phi}(l)}\right\rbrack =&
V_{1}\left( {U_{1}{\mathit{\cos}^{2}\theta} + U_{2}{\mathit{\cos}^{2}\phi}} \right) + V_{2}\left( {U_{2}{\mathit{\sin}^{2}\theta} + U_{1}{\mathit{\sin}^{2}\phi}} \right) \nonumber\\
&+ 2\xi\left( {U_{1}{\mathit{\sin}\phi}{\mathit{\cos}\theta} + U_{2}{\mathit{\cos}\phi}{\mathit{\sin}\theta}} \right) , \label{eq:var_dif_X1_Y1}\\
4U_1U_2\cdot\mathrm{Var}\left\lbrack{X_{2}^{\theta + \frac{\pi}{2}}(0) + Y_{2}^{\phi + \frac{\pi}{2}}(l)}\right\rbrack =&V_{1}\left( {U_{1}{\mathit{\sin}^{2}\theta} + U_{2}{\mathit{\sin}^{2}\phi}} \right) + V_{2}\left( {U_{2}{\mathit{\cos}^{2}\theta} + U_{1}{\mathit{\cos}^{2}\phi}} \right) \nonumber\\
&+ 2\xi\left( {U_{1}{\mathit{\cos}\phi}{\mathit{\sin}\theta} + U_{2}{\mathit{\sin}\phi}{\mathit{\cos}\theta}} \right), \label{eq:var_sum_X2_Y2} \\
4U_1U_2\cdot\mathrm{Var}\left\lbrack{X_{1}^{\theta}(0) + Y_{1}^{\phi}(l)}\right\rbrack=& V_{1}\left( {U_{1}{\mathit{\cos}^{2}\theta} + U_{2}{\mathit{\cos}^{2}\phi}} \right) + V_{2}\left( {U_{2}{\mathit{\sin}^{2}\theta} + U_{1}{\mathit{\sin}^{2}\phi}} \right) \nonumber\\
&- 2\xi\left( {U_{1}{\mathit{\sin}\phi}{\mathit{\cos}\theta} + U_{2}{\mathit{\cos}\phi}{\mathit{\sin}\theta}} \right), \label{eq:var_sum_X1_Y1} \\
4U_1U_2\cdot\mathrm{Var}\left\lbrack{X_{2}^{\theta + \frac{\pi}{2}}(0) - Y_{2}^{\phi + \frac{\pi}{2}}(l)}\right\rbrack=&V_{1}\left( {U_{1}{\mathit{\sin}^{2}\theta} + U_{2}{\mathit{\sin}^{2}\phi}} \right) + V_{2}\left( {U_{2}{\mathit{\cos}^{2}\theta} + U_{1}{\mathit{\cos}^{2}\phi}} \right) \nonumber\\
&- 2\xi\left( {U_{1}{\mathit{\cos}\phi}{\mathit{\sin}\theta} + U_{2}{\mathit{\sin}\phi}{\mathit{\cos}\theta}} \right), \label{eq:var_dif_X2_Y2}
\end{align}
\end{subequations}
where the involved coefficients are, respectively, $\xi=2{\sin\epsilon}{\sin(\beta L)}{\cosh\left(\frac{gl}{2}\right)}$, $U_1=\sin^{2}\left( {\beta l - \epsilon} \right)$, $U_2=\sin^{2}\left( {\beta l + \epsilon} \right)$, $V_1=e^{- gl}{\sin^{2}\epsilon} + {\sin^{2}(\beta l)}$, and $V_2=e^{gl}{\sin^{2}\epsilon} + {\sin^{2}(\beta l)}$.

With use of these above results (\ref{eq:var_dif_X1_Y1})--(\ref{eq:var_dif_X2_Y2}), we readily approach the following results:
\begin{subequations}
\begin{align}
\mathrm{ET}_1&=\frac{{\sin^{2}\left( {\beta l - \epsilon} \right)} + {\sin^{2}\left( {\beta l + \epsilon} \right)}}{2{\sin^{2}\left( {\beta l - \epsilon} \right)}{\sin^{2}\left( {\beta l + \epsilon} \right)}}\left\lbrack {{\cosh{(gl)}}{{\sin}^{2}\epsilon} + {{\sin}^{2}(\beta l)} + 2{\cosh\left(\frac{gl}{2}\right)}{\sin(\beta l)}{\sin\epsilon}{{\sin}\left( {\theta + \phi} \right)}} \right\rbrack, \label{eq:detail_ET1} \\
\mathrm{ET}_2&=\frac{{\sin^{2}\left( {\beta l - \epsilon} \right)} + {\sin^{2}\left( {\beta l + \epsilon} \right)}}{2{\sin^{2}\left( {\beta l - \epsilon} \right)}{\sin^{2}\left( {\beta l + \epsilon} \right)}}\left\lbrack {{\cosh(gl)}{{\sin}^{2}\epsilon} + {\mathit{\sin}^{2}(\beta l)} - 2{\cosh\left(\frac{gl}{2}\right)}{\sin(\beta l)}{\sin\epsilon}{\mathit{\sin}\left( {\theta + \phi} \right)}} \right\rbrack. \label{eq:detail_ET2}
\end{align}
\end{subequations}
Please note that compared to traditional studies, one apparent difference arises from the terms involving the argument $\beta l$. It is this aspect that piques our interest in exploring the relationship between PT symmetry and EPR correlations. Given the complexity of these formulas, we opt to utilize numerical simulations to seek the connection between the two seemingly non-overlapping research areas. 

\begin{figure}[htbp]
\centering
\fbox{\includegraphics[width=0.7\linewidth]{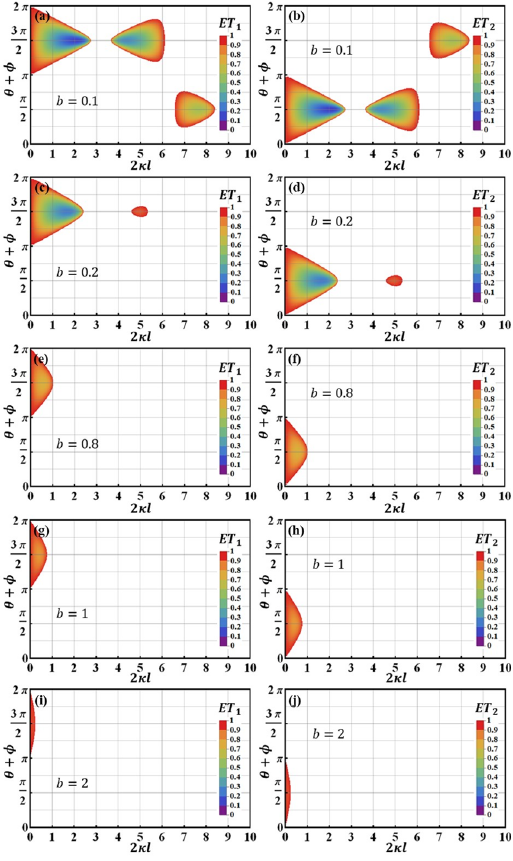}}
\caption{PT-manifested EPR stronger and weaker criteria for different $b$ values as a function of $\theta+\phi$ and $2\kappa l$. The white color signifies that the values of ET$_1$ and ET$_2$ exceed unity.}
\label{fig:two mode entanglement}
\end{figure} 

In Fig.~\ref{fig:two mode entanglement}, we present some representative simulations of ET$_1$ (\ref{eq:detail_ET1}) and ET$_2$ (\ref{eq:detail_ET2}) under varying values of $b$. Observations from these plots reveal the following: (i) Both ET$_1$ and ET$_2$ lose EPR correlations quickly with increasing $b$. (ii) Due to the similarity between the two expressions, ET$_1$ displays complementary features to ET$_2$ along the ($\theta+\phi$)-dimension. (iii) Strong EPR criteria can be approached only when $b$ is below 0.45, indicating the emergence of quantum entanglement in the system. (iv) In the regime where $b$ is small and within the PT-phase intact regime, both ET$_1$ and ET$_2$ exhibit rapidly decaying intervals of periodic strong and weak EPR correlations. Also, the simulations indicate that the competition between imbalanced amplification and the FWM process can negatively impact the EPR strong and weak criteria. This effect is especially notable when $b$ becomes large, even before the symmetry breaks down. Nevertheless, for $b$ values below 0.45, quadrature PT symmetry prevails over the FWM process, propelling the system towards the optimal quantum domain in advance. From these observations, we are aware that the dynamical participation of PSA yields distinctive effects compared to stationary participation in traditional squeezing research. Moreover, the underlying physics manifests in different forms as well. In essence, when $b$ is small, quadrature PT symmetry significantly enhances nonclassical correlations and quantum entanglement within the twin-beam system. Although there exists a trade-off between the two competing processes, notably, these discoveries represent novel insights that have not been uncovered in prior studies within the realms of NH quantum physics and traditional quantum squeezing research. 

At the EP, Eqs.~(\ref{eq:detail_ET1}) and (\ref{eq:detail_ET2}) reduce to be
\begin{subequations}
\begin{align}
\mathrm{ET}_{1}(b=1)&=\frac{e^{- 2\kappa l}( {\kappa^{2}l^{2} + 1} )\left\lbrack {1 + e^{4\kappa l} + 2\kappa^{2}l^{2}e^{2\kappa l} + 2\kappa le^{\kappa l}\left( {e^{2\kappa l} + 1} \right){\sin\left( {\theta + \phi} \right)}} \right\rbrack}{2\left( {\kappa^{2}l^{2} - 1} \right)^{2}}, \label{eq:EP_ET1} \\
\mathrm{ET}_{2}(b=1)&=\frac{e^{- 2\kappa l}( {\kappa^{2}l^{2} + 1})\left\lbrack {1 + e^{4\kappa l} + 2\kappa^{2}l^{2}e^{2\kappa l} - 2\kappa le^{\kappa l}\left( {e^{2\kappa l} + 1} \right){\sin\left( {\theta + \phi} \right)}} \right\rbrack}{2\left( {\kappa^{2}l^{2} - 1} \right)^{2}}, \label{eq:EP_ET2}
\end{align}
\end{subequations}
which have been illustrated in Figs.~\ref{fig:two mode entanglement}(g) and (h) respectively. As one can see, very weak EPR correlations persist only within a narrow range of interaction lengths, due to the high competition between PT symmetry and FWM.

Based on both expressions (\ref{eq:detail_ET1}) and (\ref{eq:detail_ET2}), as well as the simulations given in Fig.~\ref{fig:two mode entanglement}, we expect that their optimal EPR quantum correlations are associated with $\theta + \phi = \frac{3\pi}{2}$ and $\theta + \phi = \frac{\pi}{2}$, respectively. In such circumstances, ET$_{1}$ and ET$_{2}$ assume the following characteristics:
\begin{align}
\mathrm{ET}_{1}\left(\theta + \phi = \frac{3\pi}{2}\right) &=\mathrm{ET}_{2}\left(\theta + \phi = \frac{\pi}{2}\right) \nonumber\\
&= \left[ {\frac{1}{4\sin^2(\beta l - \epsilon)} + \frac{1}{4\sin^2(\beta l + \epsilon)}}\right]\left\{\left[e^{\frac{- gl}{2}}\sin\epsilon - \sin(\beta l)\right]^{2} + \left[e^{\frac{gl}{2}}\sin\epsilon -\sin(\beta l) \right]^{2} \right\}, \label{eq:best_phase_ET1}
\end{align}
a value always greater than 0. According to Cauchy–Schwartz inequality, Eq.~(\ref{eq:best_phase_ET1}) is bounded from below, yielding
\begin{align}
\mathrm{ET}_{1}\left(\theta + \phi = \frac{3\pi}{2}\right)=\mathrm{ET}_{2}\left(\theta + \phi = \frac{\pi}{2}\right)
\geq \left[{\frac{e^{\frac{- gl}{2}}{\sin\epsilon} - {\sin(\beta l)}}{2{\sin\left( {\beta l - \epsilon} \right)}} + \frac{e^{\frac{gl}{2}}{\sin\epsilon} - {\sin(\beta l)}}{2{\sin\left( {\beta l + \epsilon} \right)}}} \right]^{2}. \label{eq:inequality}
\end{align}
The equality holds if and only if the following condition is met: $[e^{\frac{gl}{2}}{\sin\epsilon} - {\sin(\beta l)}]\sin(\beta l+\epsilon)=[e^{\frac{-gl}{2}}\sin\epsilon-\sin(\beta l)]\sin(\beta l-\epsilon)$.

\subsection{Calculations of the logarithmic negativity}
In Sections 2.A and 2.B above, we have examined quantum correlations by focusing on specific properties of the system. However, it is important to note that there are various methodologies to characterize the nonclassicality properties based on the entire system. One such approach is the logarithmic negativity $E_N$ \cite{EN1,EN2,EN3,EN4}.

In this subsection, we will outline the main procedure for calculating $E_N$ as given in the main text. To this end, our first step is to compute the covariance matrix $V_{Q}$, which encapsulates the system's properties as a whole,
\begin{align}
V_{Q} = \begin{bmatrix}
\begin{matrix}
\mathrm{Var}[q_i(0)] & 0 & 0 & \mathrm{CoVar}[q_i(0),p_s(l)]\\
0 & \mathrm{Var}[p_i(0)] & \mathrm{CoVar}[p_i(0),q_s(l)] & 0\\
0 & \mathrm{CoVar}[p_i(0),q_s(l)] & \mathrm{Var}[q_s(l)] & 0\\
\mathrm{CoVar}[q_i(0),p_s(l)] & 0 & 0 & \mathrm{Var}[p_s(l)]
\end{matrix}
\end{bmatrix}. \label{eq:V_Q}
\end{align}
In Eq.~(\ref{eq:V_Q}), we have introduced the following notions:$\mathrm{Var}[x]=\langle\Delta x^2\rangle$ and $\mathrm{CoVar}[x,y]=\langle xy\rangle-\langle x\rangle\langle y\rangle$, to simplify the expression. Moreover, $V_Q$ is constructed for the column vector $\hat{V} = \left( {q_{i}(0),p_{i}(0),q_{s}(l),p_{s}(l)} \right)^{\top}$.

With use of the covariance matrix $V_Q$ (\ref{eq:V_Q}), our next step is to determine $E_N=\textrm{max}[0,-\ln{4\eta}]$ by finding out $\eta$, which is defined as follows:
\begin{equation}
\eta= \sqrt{\frac{\Sigma - \sqrt{\Sigma^{2} - 4{\det V_{Q}}}}{2}}. \label{eq:etadefination} 
\end{equation}

In order to evaluate $\eta$, one needs to work out $\Sigma$, which is given by
\begin{equation}
\Sigma = \mathrm{Var}[q_{i}(0)] \mathrm{Var}[p_{i}(0)] + \mathrm{Var}[q_{s}(l)]\mathrm{Var}[p_{s}(l)] + 2\mathrm{CoVar}[q_{i}(0),p_{s}(l)]\mathrm{CoVar}[p_{i}(0),q_{s}(l)].\label{eq:definitionSigma}
\end{equation}
After some lengthy derivation, we eventually find $\eta$ to be
\begin{equation}
\frac{\sqrt{[\beta^2 + \kappa^2\sin^2(\beta l)]^2 + 4\big\{\beta^2\kappa^2\sin^2(\beta l)\cosh(gl) - \kappa\cosh(\frac{gl}{2})\sqrt{\beta^2\sin^2(\beta l)\left[2\beta^2\kappa^2\sin^2(\beta l)\cosh(gl) + \kappa^4\sin^4(\beta l) + \beta^4 \right]} \big\}}}{{|g^2 - 4\kappa^2\cos^2(\beta l)|}}. \label{eq:eta}
\end{equation}
From the preceding discussions, we have learnt that in both type-I and type-II quadrature PT symmetry, the sharp change in quantum correlation typically does not align precisely with $b=1$ (i.e., the EP). Nonetheless, it is good to have the expression of $\eta$ at the EP, which takes the form of
\begin{align}
\eta_{EP} = \sqrt{\frac{\left( {\kappa^{2}l^{2} + 1} \right)^{2} + 4\left\lbrack {\kappa^{2}l^{2}{\cosh(2\kappa l)} - \kappa l{\cosh(\kappa l)}\sqrt{1 + \kappa^{4}l^{4} + 2\kappa^{2}l^{2}{\cosh(2\kappa l)}}} \right\rbrack}{16\left( {\kappa^{2}l^{2} - 1} \right)^{2}}}. \label{eq:eta_EP}
\end{align}
The derived results above provide the foundation for the plot of the logarithmic negativity $E_N$ shown in Fig.~3(b) of the main text across the PT phase transition for various $b$ values.

Before proceeding to the next session, it is noteworthy to mention that the discussions presented in this section regarding the exploration of the relationship between PT symmetry and quantum correlations/entanglement can be extended to the type-I case, with exception of its complexity stemming from the Langevin noise in theory.

\section{Further discussion on quantum sensing in type-II quadrature PT symmetry}
As pointed out in the main text, a fundamental distinction between type-I and type-II in quantum sensing lies in the fact that the latter scheme explicitly demonstrates enhanced quantum sensitivity in the PT-unbroken region, inaccessible to the former. This improvement is achieved through the combined effect of PT symmetry and quantum squeezing. However, the presence of Langenvin noise can quickly degrade this quantum sensitivity, pushing the system into the classical domain, as shown in our earlier work on type-I quadrature PT symmetry \cite{1}.

In this section, we will delve deeper into our calculations of quantum sensing, expanding upon the points briefly mentioned in the main text. Our focus will be on providing more comprehensive details regarding the theoretical derivations and conducting further analysis on the quantum sensing performance, particularly in the type-II scenario with an initial two-photon coherent state $|\alpha,\alpha\rangle$ as an input.

\subsection{Calculations of the quantum Fisher information $F_{\kappa}$}
\begin{figure}[htbp]
\centering
\fbox{\includegraphics[width=0.8\linewidth]{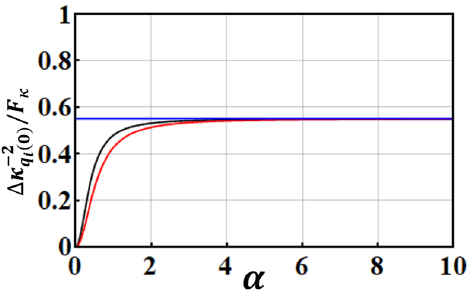}}
\caption{Plot of $\Delta\kappa^{-2}{q_i(0)}/F\kappa$ as a function of $\alpha$ for $l=2\pi/\beta$, with parameters ${b=0.2,\kappa=0.5}$. The red (black) line represents $F_\kappa$ calculated using the covariance matrix (final state vector $\vert\psi_f\rangle$). The blue line corresponds to $F_{\kappa}$ obtained by excluding the first term on the right-hand side (i.e., the second line of Eq.~(\ref{eq:F_kappa})).}
\label{fig:xxx}
\end{figure}

For Gaussian states \cite{gaussian} (e.g., our case), the quantum Fisher information can be computed in the following way,
\begin{align}
F_{\kappa}\approx&\frac{1}{2}\text{Tr}\left(V_{out}^{- 1}\frac{d V_{out}}{d\kappa}V_{out}^{- 1}\frac{d V_{out}}{d\kappa}\right)+ \left( \frac{d\mu_{out}}{d\kappa} \right)^{\top}V_{out}^{- 1}\frac{d\mu_{out}}{d\kappa}\nonumber\\
\approx&\left( \frac{d\mu_{out}}{d\kappa} \right)^{\top}V_{out}^{- 1}\frac{d\mu_{out}}{d\kappa} \;(\text{when}\;\alpha\;\text{is large}).
\label{eq:F_kappa}
\end{align}
In Eq.~(\ref{eq:F_kappa}), the amplitude vector $\mu_{out}$ is defined in the quadrature basis via $\mu_{j} = \langle {\hat{v}}_{j} \rangle$ and $V_{j,k} = \frac{1}{2}\langle {{\hat{v}}_{j}{\hat{v}}_{k} + {\hat{v}}_{k}{\hat{v}}_{j}} \rangle - \langle {\hat{v}}_{j} \rangle\langle {\hat{v}}_{k}\rangle$, for $1 \leq j,k \leq 2$, with the column vector $\hat{v} = \left( {q_{i}(0),q_{s}(l),p_{i}(0),p_{s}(l)} \right)^{\top}$. It is easy to check that here $V_{out}$ resembles the covariance matrix $V_Q$ described in Eq.~(\ref{eq:V_Q}), with the only difference being the interchange of the second and third rows. In general, the quantum Fisher information $F_{\kappa}$ (\ref{eq:F_kappa}) characterizes the overall SNR of the whole system. Furthermore, the column vector $\frac{d\mu_{out}}{d\kappa}$ assumes the following form,
\begin{align}
\frac{d\mu_{out}}{d\kappa} = \left\lbrack {\chi_{\kappa}^{q_{i}{(0)}},\chi_{\kappa}^{q_{s}{(l)}},\chi_{\kappa}^{p_{i}{(0)}},\chi_{\kappa}^{p_{s}{(l)}}} \right\rbrack^{\top}, \label{eq:muout}
\end{align}
with $\chi_w^A\equiv\frac{\partial\langle A\rangle}{\partial w}$ being the susceptibility, a quantity characterizing the system's response to the parameter $w$ of interest by measuring the quantum observable $A$. One important point to note regarding Eq.~(\ref{eq:F_kappa}) in the first line is that as $\alpha$ rises, the significance of the first term on the right-hand side diminishes. To illustrate, in Fig.~\ref{fig:xxx}, we have conducted a comparison, showing that when $\alpha>3$, the impact of the first term becomes negligible.

The final neat expression of $F_{\kappa}$ is found to be
\begin{eqnarray}
F_{\kappa}\approx\frac{4\left\{[\cosh(gl)+1][g^2\sin(\beta l)-4\kappa^2\beta l\cos(\beta l)]^2+2g^2\kappa^2[\sin(2\beta l)-2\beta l]^2\right\}}{\beta^2[g^2-4\kappa^2\cos(\beta l)^2]^2}\nonumber\\+\frac{16\alpha^2\left\{\cosh(gl)[g^2\sin(\beta l)-4\kappa^2\beta l\cos(\beta l)]^2+g^2\kappa^2[\sin(2\beta l)-2\beta l]^2\right\}}{\beta^2[g^2-4\kappa^2\cos(\beta l)^2]^2},\label{eq:fk}
\end{eqnarray}
which amounts to the quantum Fisher information of the entire system, so it aligns with all other quantum observables including single-mode and two-mode quadratures as well as relative-intensity squeezing. Equation~(\ref{eq:fk}) is highly complex and lacks intuitiveness. Therefore, we resort to numerical simulations in this context to illustrate its behavior. For different values of $b$, one can visualize how $F_{\kappa}$ behaves from Fig.~4 in the main text and the accompanying Figs.~\ref{fig:inverse_variances_F_k}, \ref{fig:near_EP_inverse_variances_F_k}, and \ref{fig:two_mode_inverse_variances_F_k} below. While $F_{\kappa}$ may not hold significance in the vicinity of the EP (i.e., when $g\rightarrow 2\kappa$), having its reduced format available can still provide valuable insights,
\begin{equation}
F_{\kappa}(g\rightarrow2\kappa) \approx \frac{4l^{2}\left\lbrack {9\kappa^{4}l^{4}-6\kappa^{2}l^{2}+9+\left( {\kappa^{2}l^{2} - 3} \right)^{2}{\cosh(2\kappa L)}} \right\rbrack+16\alpha^{2}l^{2}\left\lbrack {4\kappa^{4}l^{4} + \left( {\kappa^{2}l^{2} - 3} \right)^{2}{\cosh(2\kappa L)}} \right\rbrack}{9\left( {\kappa^{2}l^{2} - 1} \right)^{2}}.\label{eq:FK_EP}
\end{equation}
In Eq.~(\ref{eq:FK_EP}), it is observed that $F_{\kappa}$ becomes divergent (i.e., infinite) when $\kappa l=1$, which poses disadvantage for quantum sensing purposes. Put differently, utilizing type-II quadrature PT symmetry for enhancing sensitivity at or near EPs seems impractical. We will get back to this point shortly.

Here, as an additional note, it's worth mentioning that, for the current system, the quantum Fisher information can also be computed by
\begin{equation}
F_{\kappa}=4(\langle\partial_{\kappa}\psi_f|\partial_{\kappa}\psi_f\rangle-|\langle\partial_{\kappa}\psi_f|\psi_f\rangle|^2),\nonumber
\end{equation}
for the final system state $\vert\psi_f\rangle$ in the Schr\"{o}dinger representation. The final state evolves as $|\psi_f\rangle=\widehat{D}\left[\alpha_i(0),\alpha_s(l)\right]|\mathrm{TMCSV}\rangle$ with $|\mathrm{TMCSV}\rangle=S^{(2)}|0,0\rangle$ being the two-mode combination squeezed vacuum state and $S^{(2)}$ the two-mode squeezing operator. Next, we can cast the above $F_{\kappa}$ as 
\begin{equation}
F_{\kappa}=4\left(\langle\mathrm{TMCSV}\left|\hat{\xi}^{\dagger} \hat{\xi}\right|\mathrm{TMCSV}\rangle-\langle\mathrm{TMCSV}\left|\hat{\xi}^{\dagger}\right|\mathrm{TMCSV}\rangle\langle\mathrm{TMCSV}|\hat{\xi}|\mathrm{TMCSV}\rangle\right),\nonumber
\end{equation}
where $\hat{\xi}=\partial_{\kappa}-\frac{\left[\vert\partial_{\kappa}\alpha_i(0)\vert^2+\vert\partial_{\kappa}\alpha_s(l)\vert^2\right]}{2}-\left[a_i(l)\partial_{\kappa}\alpha^{*}_i(0)+a_s(0)\partial_{\kappa}\alpha^{*}_s(l)\right]+\left[a^{\dagger}_i(l)+\alpha^{*}_i(0)\right]\partial_{\kappa}\alpha_i(0)+\left[a^{\dagger}_s(0)+\alpha^{*}_s(l)\right]\partial_{\kappa}\alpha_s(l)$. 

We can prove that when $\alpha$ becomes large, the above two methods become equivalent for type-II quadrature PT systems, producing identical results for $F_{\kappa}$ as referenced in Eq.~(\ref{eq:fk}). To have a visual presentation, in Fig.~\ref{fig:xxx}, we have graphed the ratio of $\Delta\kappa^{-2}_{q_i(0)}/F_{\kappa}$ for $l=2\pi/\beta$, with $F_{\kappa}$ computed using the aforementioned two methods. As one can see from the figure, it becomes evident that the two methods converge to similar values when $\alpha>2$. However, this equivalence does not extend to type-I systems owing to the interference of Langevin noise. In practice, the calculation of $F_{\kappa}$ using the covariance matrix offers broader applicability.

\subsection{Calculations of susceptibilities}
Following the definition of the susceptibility introduced in the main text, we can readily derive the system response, i.e., the susceptibility $\chi$, to a tiny perturbation on the parameter $\kappa$ based on measuring either of the single-mode quadratures $\left\lbrack {q_{s}(l),p_{s}(l),q_{i}(0),p_{i}(0)} \right\rbrack$. For the sake of simplicity, we will hereafter assume that the average photon number of the thermal bosonic mode from the environment is zero.

In order to compute the $\chi$, we first need to find out the expectation or mean value of each quadrature when seeding a two-photon coherent state $\left| \psi_{i} \right\rangle  = \left|{\alpha,\alpha} \right\rangle$. After performing some calculations, we get
\begin{subequations}
\begin{align}
\left\langle {q_{i}(0)} \right\rangle &= e^{- \frac{gl}{2}}\csc\left( {\beta l + \epsilon} \right){\mathit{\sin}\epsilon}\left\langle {q_{i}(l)} \right\rangle - \csc\left( {\beta l + \epsilon} \right){\mathit{\sin}(\beta l)}\left\langle {p_{s}(0)} \right\rangle, \label{eq:qi_expectedvalue}\\
\left\langle {p_{i}(0)} \right\rangle &= e^{\frac{gl}{2}}\csc\left( {\beta l - \epsilon} \right){\mathit{\sin}\epsilon}\left\langle {p_{i}(l)} \right\rangle - \csc\left( {\beta l - \epsilon} \right){\mathit{\sin}(\beta l)}\left\langle {q_{s}(0)} \right\rangle, \label{eq:pi_expectedvalue}\\
\left\langle {q_{s}(l)} \right\rangle &= e^{- \frac{gl}{2}}\csc\left( {\beta l - \epsilon} \right){\mathit{\sin}\epsilon}\left\langle {q_{s}(0)} \right\rangle - \csc\left( {\beta l - \epsilon} \right){\mathit{\sin}(\beta l)}\left\langle {p_{i}(l)} \right\rangle, \label{eq:qs_expectedvalue}\\
\left\langle {p_{s}(l)} \right\rangle &= e^{\frac{gl}{2}}\csc\left( {\beta l + \epsilon} \right){\mathit{\sin}\epsilon}\left\langle {p_{s}(0)} \right\rangle - \csc\left( {\beta l + \epsilon} \right){\mathit{\sin}(\beta l)}\left\langle {q_{i}(l)} \right\rangle. \label{eq:ps_expectedvalue}
\end{align}
\end{subequations}
Using the results from Eqs.~(\ref{eq:qi_expectedvalue}) to (\ref{eq:ps_expectedvalue}), we can readily calculate the four susceptibilities, which are
\begin{subequations}
\begin{align}
\chi_{\kappa}^{q_{i}{(0)}} &= \frac{{e}^{- \frac{gl}{2}}{\mathit{\sin}(\beta l)}\left\lbrack {2\kappa l + \left( {2 - gl} \right){\mathit{\cos}\epsilon}} \right\rbrack - 2\beta l{e}^{- \frac{gl}{2}}{\mathit{\cos}\epsilon}{\mathit{\cos}(\beta l)} - {\mathit{\sin}\epsilon}\left( {2\kappa l + {\mathit{\cos}\epsilon}} \right) + {\mathit{\cos}\epsilon}{\mathit{\sin}\left( {2\beta l + \epsilon} \right)}}{2\beta~{\sin}^{2}\left( {\beta l + \epsilon} \right)}\alpha, \label{eq:qixk}\\
\chi_{\kappa}^{p_{i}{(0)}} & = \frac{2\beta l{e}^{\frac{gl}{2}}{\mathit{\cos}\epsilon}{\mathit{\cos}(\beta l)} - {e}^{\frac{gL}{2}}{\mathit{\sin}(\beta l)}\left\lbrack {\left( {2 + gl} \right){\mathit{\cos}\epsilon} - 2\kappa l} \right\rbrack + {\mathit{\sin}\epsilon}\left( {{\mathit{\cos}\epsilon} - 2\kappa l} \right) + {\mathit{\cos}\epsilon}{\mathit{\sin}\left( {2\beta l + \epsilon} \right)}}{2\beta~{\sin}^{2}\left( {\beta l - \epsilon} \right)}\alpha, \label{eq:pixk}\\
\chi_{\kappa}^{q_{s}{(l)}} & = \frac{2\beta l{e}^{- \frac{gl}{2}}{\mathit{\cos}\epsilon}{\mathit{\cos}(\beta l)} - {e}^{- \frac{gl}{2}}{\mathit{\sin}(\beta l)}\left\lbrack {\left( {2 + gl} \right){\mathit{\cos}\epsilon} - 2\kappa l} \right\rbrack + {\mathit{\sin}\epsilon}\left( {{\mathit{\cos}\epsilon} - 2\kappa l} \right) + {\mathit{\cos}\epsilon}{\mathit{\sin}\left( {2\beta l + \epsilon} \right)}}{2\beta~{\sin}^{2}\left( {\beta l - \epsilon} \right)}\alpha, \label{eq:qsxk}\\
\chi_{\kappa}^{p_{s}{(l)}} &= \frac{{e}^{\frac{gl}{2}}{\mathit{\sin}(\beta l)}\left\lbrack {2\kappa l + \left( {2 - gl} \right){\mathit{\cos}\epsilon}} \right\rbrack - 2\beta l{e}^{\frac{gl}{2}}{\mathit{\cos}\epsilon}{\mathit{\cos}(\beta l)} - {\mathit{\sin}\epsilon}\left( {2\kappa l + {\mathit{\cos}\epsilon}} \right) + {\mathit{\cos}\epsilon}{\mathit{\sin}\left( {2\beta l + \epsilon} \right)}}{2\beta~{\sin}^{2}\left( {\beta l + \epsilon} \right)}\alpha. \label{eq:psxk} 
\end{align}
\end{subequations}

\begin{figure}[htbp]
\centering
\fbox{\includegraphics[width=0.95\linewidth]{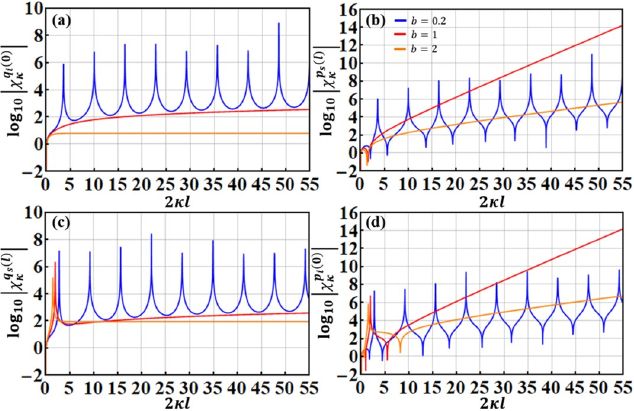}}
\caption{The logarithmic susceptibilities $\chi_{\kappa}^{q_{i}{(0)}}$ (a), $\chi_{\kappa}^{p_{s}{(l)}}$ (b), $\chi_{\kappa}^{q_{s}{(l)}}$ (c), and $\chi_{\kappa}^{p_{i}{(0)}}$ (d) plotted for the parameters $\{\alpha=10$ and $\kappa=0.5\}$ across various $b$ values under the measurement of a single-mode quadrature $\{q_i(0),p_s(l),q_s(l),p_i(0)\}$.}
\label{fig:susceptibility}
\end{figure} 

To have an intuitive understanding of the behaviors of these susceptibilities (\ref{eq:qixk})--(\ref{eq:psxk}), we have numerically plotted them in Figs.~\ref{fig:susceptibility}(a)--(d) respectively. As one can see, in the PT-phase unbroken region ($b<1$, indicated by blue), all four susceptibilities exhibit periodic fluctuations with a period $T=2\pi\kappa/\beta$. Notably, $\chi_{\kappa}^{q_{i}{(0)}}$ and $\chi_{\kappa}^{p_{s}{(l)}}$ peak at locations $(nT-2\epsilon\kappa/\beta)$, while $\chi_{\kappa}^{q_{s}{(l)}}$ and $\chi_{\kappa}^{p_{i}{(0)}}$ peak at locations $(nT+2\epsilon\kappa/\beta)$, aligning with the peak positions of corresponding quadrature variances as discussed in the main text. These susceptibility peaks indicate particular strong response signals, implying the (sub)optimal locations for superior sensing measurement. 

Oppositely, in the phase-broken regime ($b>1$), the periodic patterns vanish. Specifically, $\chi^{q_i(0)}_{\kappa}$ displays a smooth, monotonically increasing curve bounded above by the $b=1$ curve. $\chi^{p_s(l)}_{\kappa}$ shows a dip initially, followed by monotonic increase with an upper bound from the $b=1$ curve. In contrast, $\chi^{q_s(l)}_{\kappa}$ demonstrates a solitary peak near the beginning, contrasting with $\chi^{p_s(l)}_{\kappa}$. The pattern of $\chi^{p_i(0)}_{\kappa}$ becomes a bit complicated; starting with a dip, followed by a peak, a secondary dip, and then monotonic growth. 

Noteworthy is that for all four susceptibilities, as $\kappa l$ increases, they all grow monotonically, bounded above by the $b=1$ curve, regardless of passing through a dip or a peak.

\begin{figure}[htbp]
\centering
\fbox{\includegraphics[width=0.95\linewidth]{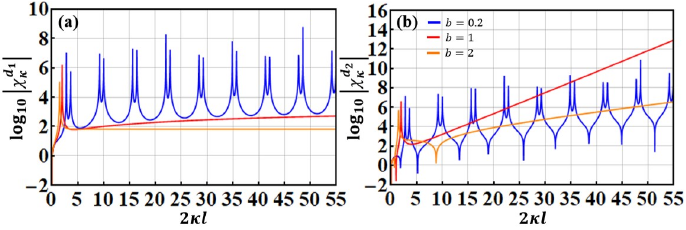}}
\caption{The logarithmic susceptibilities $\chi_{\kappa}^{d_{1}}$ (a) and $\chi_{\kappa}^{d_{2}}$ (b) for the parameters $\{\alpha=10$ and $\kappa=0.5\}$ and different $b$ values under the measurement of a two-mode quadrature $\{d_1,d_2\}$.}
\label{fig:two_mode_susceptibility}
\end{figure} 

On the other hand, we can also analyze the system's response, i.e., the susceptibility, by measuring two-mode quadratures. It can be demonstrated that the corresponding susceptibilities are now 
\begin{subequations}
\begin{align}
\chi_{\kappa}^{d_{1}} &= \frac{\partial(\left\langle {q_{i}(0)} \right\rangle+\left\langle {q_{s}(l)} \right\rangle)}{\sqrt{2}\partial\kappa}=\frac{\chi_{\kappa}^{q_{i}{(0)}}+\chi_{\kappa}^{q_{s}{(l)}}}{\sqrt{2}}, \label{eq:d1xk}\\
\chi_{\kappa}^{d_{2}} &= \frac{\partial(\left\langle {p_{i}(0)} \right\rangle+\left\langle {p_{s}(l)} \right\rangle)}{\sqrt{2}\partial\kappa}=\frac{\chi_{\kappa}^{p_{i}{(0)}}+\chi_{\kappa}^{p_{s}{(l)}}}{\sqrt{2}},\label{eq:d2xk}
\end{align}
\end{subequations}
which are essentially combinations of susceptibilities (\ref{eq:qixk})--(\ref{eq:psxk}) derived from single-mode quadratures. Using the same parameters as in Fig.~\ref{fig:susceptibility}, the behaviors of $\chi_{\kappa}^{d_{1}}$ and $\chi_{\kappa}^{d_{2}}$ are presented in Figs.~\ref{fig:two_mode_susceptibility}(a) and (b) respectively. Comparing with Fig.~\ref{fig:susceptibility}, $\chi^{d_1}_{\kappa}$ shows periodic bimodal oscillations before the phase transition and a sole peak after the transition, whereas $\chi^{d_1}_{\kappa}$ showcases periodic dips followed by double peaks in the unbroken phase region and a dip followed by a peak and a secondary dip in the broken region. It is worth noting that the $b=1$ curve in the $\chi^{d_1}_{\kappa}$ case only exhibits a dip followed by a single peak.

All in all, a sharp peak in $\chi$ indicates that the system responds sharply around that location to an infinitesimal perturbation in the observed quantity. Contrarily, a flat curve suggests that the system lacks a sensitive response to a small perturbation of the observed quantity.

\subsection{Calculations of inverse variances}
In order to assess the system's sensing performance using the quantum Fisher information $F_{\kappa}$, it is essential to juxtapose it with the inverse variances. Here, in line with our previous discussion on susceptibilities (Section 3.B), we will also look at two cases: single-mode quadratures versus two-mode quadratures. As demonstrated below, varying quantum measurement techniques applied to the same system can result in differing sensitivity levels and signal-to-noise (SNR) ratios. Specifically, our current study reveals that the two-mode quadrature surpasses the single-mode quadrature in sensing performance in general.
 
\emph{\textcolor{blue}{Single-Mode Quadrature Case}}---The inverse variance, critical for determining the ultimate precision of the parameter estimation for $\kappa$, is defined as ${\Delta\kappa}^{-2}\equiv(\chi^O_{\kappa})^2/\langle\Delta O^2\rangle$, where $O$ represents the observable for either of the four single-mode quadratures. For the single-mode quadrature case, specifically, comparing $\Delta\kappa^{-2}_{p_i(0)}$, $\Delta\kappa^{-2}_{q_i(0)}$, $\Delta\kappa^{-2}_{q_s(l)}$, and $\Delta\kappa^{-2}_{p_s(l)}$ with $F_{\kappa}$ becomes crucial.  These inverse variances relative to the quantum Fisher information must satisfy the following inequalities: $F_{\kappa} \geq \left\{ {{{\Delta}\kappa}_{q_s(l)}^{- 2},{{\Delta}\kappa}_{p_s(l)}^{-2},{{\Delta}\kappa}_{q_i(0)}^{-2},{{\Delta}\kappa}_{p_i(0)}^{-2}} \right\}$. In other words, the Cram\'{e}r-Rao lower bound is determined by $F^{-1}_{\kappa}\Delta\kappa^{-2}\leq1$. (In the insets of Fig.~4 in the main text, we have illustrated such a ratio for each single-mode quadrature for your reference. A ratio approaching unity indicates the location where the best sensitivity is available. However, whether the nature of sensing is classical or quantum depends on whether the observable's nature at those locations is classical or quantum.)
 
\begin{figure}[htbp]
\centering
\fbox{\includegraphics[width=0.9\linewidth]{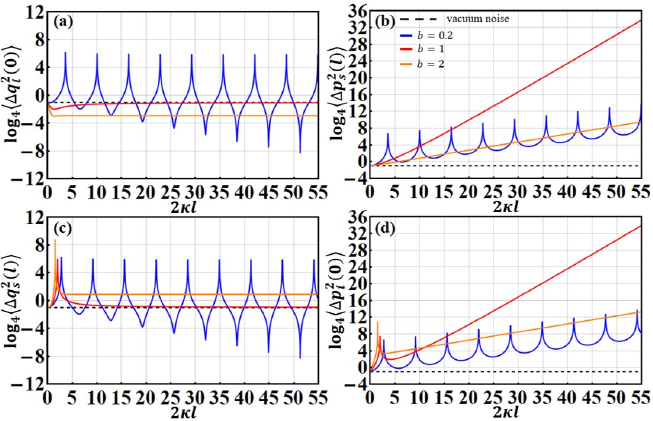}}
\caption{The logarithmic variances $\langle\Delta q^2_i(0)\rangle$ (a), $\langle\Delta p^2_s(l)\rangle$ (b), $\langle\Delta q^2_s(l)\rangle$ (c), and $\langle\Delta p^2_i(0)\rangle$ (d) for the parameter $\kappa=0.5$ and different $b$ values. The dashed lines represent the vacuum noise for comparison.}
\label{fig:variances_susceptibility}
\end{figure}

 As such, our next step is to determine these inverse variances and conduct comparisons with $F_{\kappa}$. To accomplish this, we initiate similar calculations to those performed for the vacuum input state, this time recalculating the variances for the two-photon coherent seeding state $|\psi_i\rangle = |\alpha,\alpha\rangle$. After some necessary calculations, we reach the following results:
 \begin{subequations}
\begin{align}
\langle\Delta q^2_i(0)\rangle&=\frac{e^{-gl}\sin^2\epsilon+\sin^2(\beta l)}{4\sin^2(\beta l+\epsilon)}, \label{eq:qivar} \\
\langle\Delta p^2_i(0)\rangle&=\frac{e^{gl}\sin^2\epsilon+\sin^2(\beta l)}{4\sin^2(\beta l-\epsilon)}, \label{eq:pivar}\\
\langle\Delta q^2_s(l)\rangle&=\frac{e^{-gl}\sin^2\epsilon+\sin^2(\beta l)}{4\sin^2(\beta l-\epsilon)}, \label{eq:qsvar} \\
\langle\Delta p^2_s(l)\rangle&=\frac{e^{gl}\sin^2\epsilon+\sin^2(\beta l)}{4\sin^2(\beta l+\epsilon)}. \label{eq:psvar}
\end{align}
\end{subequations}
In Fig.~\ref{fig:variances_susceptibility}, we have numerically simulated these variances with $\kappa=0.5$ for different $b$ values. Upon comparison with the variances (3a)--(3d) of single-mode quadratures in the main text, we observe that they are in fact identical.

With these variances (\ref{eq:qivar})--(\ref{eq:psvar}) and susceptibilities (\ref{eq:qixk})--(\ref{eq:psxk}), calculating the corresponding inverse variances becomes straightforward. A simple calculation leads us to the following important results:
\begin{subequations}
\begin{align}
\Delta\kappa_{q_i(0)}^{- 2} &= \frac{\alpha^{2}\left\{ {e^{- \frac{gl}{2}}{\sin(\beta l)}\left\lbrack {2\kappa l + \left( {2 - gl} \right){\cos\epsilon}} \right\rbrack - 2\beta l{e}^{- \frac{gl}{2}}{\cos\epsilon}{\cos(\beta l)} - {\sin\epsilon}( {2\kappa l + {\cos\epsilon}}) + {\cos\epsilon}{\sin\left( {2\beta l + \epsilon} \right)}} \right\}^{2}}{\beta^{2}~{\sin}^{2}\left( {\beta l + \epsilon} \right)\left[ {e^{- gl}{\sin^{2}\epsilon} + {\sin^{2}(\beta l)}} \right]}, \label{eq:qidk}\\
\Delta\kappa_{p_i(0)}^{- 2} &= \frac{\alpha^{2}\left\{ {2\beta le^{\frac{gl}{2}}{\cos\epsilon}{\cos(\beta l)} - {e}^{\frac{gl}{2}}{\sin(\beta l)}\left\lbrack {\left( {2 + gl} \right){\cos\epsilon} - 2\kappa l} \right\rbrack + {\sin\epsilon}\left( {{\cos\epsilon} - 2\kappa l} \right) + {\cos\epsilon}{\sin\left( {2\beta l + \epsilon} \right)}} \right\}^{2}}{\beta^{2}~{\sin}^{2}\left( {\beta l - \epsilon} \right)\left[{e^{gl}{\sin^{2}\epsilon} + {\sin^{2}(\beta l)}} \right]}, \label{eq:pidk}\\
\Delta\kappa_{q_s(l)}^{- 2} &= \frac{\alpha^{2}\left\{ {2\beta le^{- \frac{gl}{2}}{\cos\epsilon}{\cos(\beta l)} - {e}^{- \frac{gl}{2}}{\sin(\beta l)}\left\lbrack {\left( {2 + gl} \right){\cos\epsilon} - 2\kappa l} \right\rbrack + {\sin\epsilon}\left( {{\cos\epsilon} - 2\kappa l} \right) + {\cos\epsilon}{\sin\left( {2\beta l + \epsilon} \right)}} \right\}^{2}}{\beta^{2}~{\sin}^{2}\left( {\beta l - \epsilon} \right)\left[{e^{- gl}{\sin^{2}\epsilon} + {\sin^{2}(\beta l)}} \right]}, \label{eq:qsdk}\\
\Delta\kappa_{p_s(l)}^{- 2} &= \frac{\alpha^{2}\left\{ {e^{\frac{gl}{2}}{\sin(\beta l)}\left\lbrack {2\kappa l + \left( {2 - gl} \right){\cos\epsilon}} \right\rbrack - 2\beta l{e}^{\frac{gl}{2}}{\cos\epsilon}{\cos(\beta l)} - {\sin\epsilon}\left( {2\kappa l + {\cos\epsilon}} \right) + {\cos\epsilon}{\sin\left( {2\beta l + \epsilon} \right)}} \right\}^{2}}{\beta^{2}~{\sin}^{2}\left( {\beta l + \epsilon} \right)\left[ {e^{gl}{\sin^{2}\epsilon} + {\sin^{2}(\beta l)}} \right]}. \label{eq:psdk} 
\end{align}
\end{subequations}

\begin{figure}[htbp]
\centering
\fbox{\includegraphics[width=0.9\linewidth]{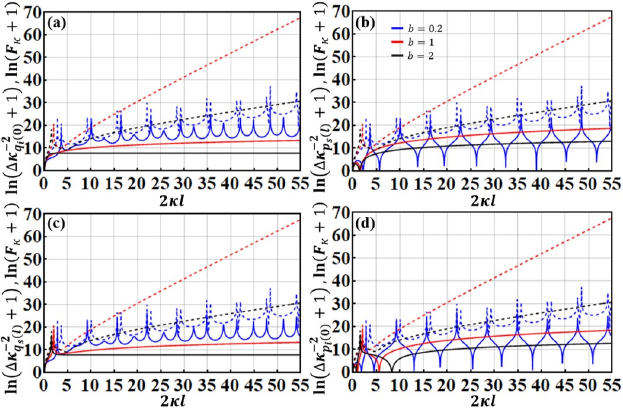}}
\caption{Comparisons between the logarithmic inverse variances $\log_{10}\left({\Delta\kappa}_{q_{i}(0)}^{- 2}+1\right)$ (a), $\log_{10}\left({\Delta\kappa}_{p_{s}(l)}^{- 2}+1\right)$ (b), $\log_{10}\left({\Delta\kappa}_{q_{s}(l)}^{- 2}+1\right)$ (c), and $\log_{10}\left({\Delta\kappa}_{p_{i}(0)}^{- 2}+1\right)$ (d) of four single-mode quadrature observables and the logarithmic quantum Fisher information $\log_{10}(F_{\kappa}+1)$ as a function of dimensionless length for parameters $\{\alpha=10,\kappa=0.5\}$ and different $b$ values. The solid lines correspond to $\log_{10}\left({\Delta\kappa}^{- 2}+1\right)$, while the dashed lines correspond to $\log_{10}(F_{\kappa}+1)$.}
\label{fig:inverse_variances_F_k}
\end{figure} 

\begin{figure}[t]
\centering
\fbox{\includegraphics[width=0.9\linewidth]{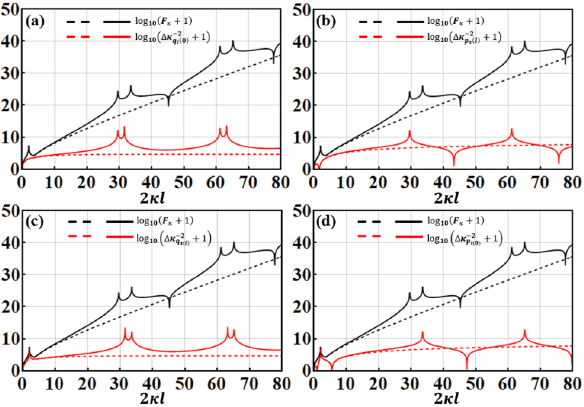}}
\caption{Comparisons between the logarithmic inverse variances $\log_{10}\left({\Delta\kappa}_{q_{i}(0)}^{- 2}+1\right)$ (a), $\log_{10}\left({\Delta\kappa}_{p_{s}(l)}^{- 2}+1\right)$ (b), $\log_{10}\left({\Delta\kappa}_{q_{s}(l)}^{- 2}+1\right)$ (c), and $\log_{10}\left({\Delta\kappa}_{p_{i}(0)}^{- 2}+1\right)$ (d) of single-mode quadrature observables and the logarithmic quantum Fisher information $\log_{10}(F_{\kappa}+1)$ as a function of dimensionless length for parameters $\{\alpha=10,\kappa=0.5\}$ and different $b$ values in the proximity of the EP. In particular, both approaches to the EP are considered, with the solid lines representing $b=0.98$ and the dashed lines representing $b=1.02$.}
\label{fig:near_EP_inverse_variances_F_k}
\end{figure} 

Despite the complexity of the mathematical expressions (\ref{eq:qidk})--(\ref{eq:psdk}), we can gain insight into the behaviors of the inverse variances $\left\{ {{{\Delta}\kappa}_{q_i(0)}^{- 2},{{\Delta}\kappa}_{p_i(0)}^{- 2},{{\Delta}\kappa}_{q_s(l)}^{- 2},{{\Delta}\kappa}_{p_s(l)}^{- 2}} \right\}$ by numerically comparing them with $F_{\kappa}$, as depicted in Fig.~\ref{fig:inverse_variances_F_k} for various $b$ values. Understanding these relationships aids our comprehension. From the definition of ${\Delta\kappa}^{-2}$, we recognize that the susceptibility $\chi^O_{\kappa}$ and the variance $\langle\Delta O^2\rangle$ jointly determine the overall measurement accuracy. Furthermore, as discussed in Section 3.B, the peak locations of $\chi^O_{\kappa}$ coincide with those of $\langle\Delta O^2\rangle$ and closely approach the $F_{\kappa}$ bound, suggesting these locations as potential optimal sensing measurement points. On the other hand, examining Fig.~\ref{fig:variances_susceptibility}, we observe that the corresponding single-mode quadrature variances exhibit classical characteristics. This observation leads to the conclusion that, in the phase unbroken region, optimal classical sensing measurements are only attainable at these peak locations.

In addition to this finding, our focus shifts to whether optimized quantum sensing measurements are achievable in the type-II scheme. By carefully examining Figs.~\ref{fig:susceptibility}(a) and (c), Figs.~\ref{fig:variances_susceptibility}(a) and (c), and Figs.~\ref{fig:inverse_variances_F_k}(a) and (c), we discern that (sub)optimal quantum sensing can indeed be achieved in the unbreaking symmetry region by measuring $q_i(0)$ or $q_s(l)$ at the dip locations ($nT$) in Figs.~\ref{fig:variances_susceptibility}. These dip locations correspond to the series of secondary small peaks in Figs.~\ref{fig:inverse_variances_F_k}(a) and (c). Notably, such accomplishment arises from they synergistic interplay of PT symmetry and quantum squeezing; either alone cannot facilitate this (sub)optimal quantum sensing. This is a stark contrast to the type-I scenario \cite{1} where quantum sensing is challenging to be reached because of the Langevin noise. However, as $b$ surpasses unity, despite the presence of quantum squeezing, achieving (sub)optimal quantum sensing becomes unfeasible. This is attributed to the substantially monotonic increase in $F_{\kappa}$ due to the rapidly expanded Hilbert space of the system's final state from the PSA. 

Our previous work on the type-I system \cite{1} revealed that there is no enhancement in quantum sensitivity in the vicinity of EP. It is now interesting to know whether this statement holds true for a type-II system as well. To address this, we conducted numerical analyses and presented the results in Fig.~\ref{fig:near_EP_inverse_variances_F_k}. As we can see from Fig.~\ref{fig:near_EP_inverse_variances_F_k}, regardless of the approach to the EP, optimal quantum sensing remains unattainable. This mirrors the findings in type-I systems \cite{1}: achieving EP-enhanced quantum sensors in PT-symmetric quantum systems proves challenging with fair sampling measurements.

\begin{figure}[htbp]
\centering
\fbox{\includegraphics[width=.9\linewidth]{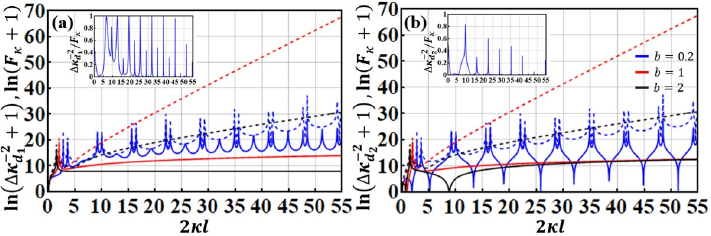}}
\caption{Comparisons between the logarithmic inverse variances $\ln\left(\Delta\kappa_{d_1}^{- 2}+1\right)$ (a) and $\ln\left(\Delta\kappa_{d_2}^{- 2}+1\right)$ (b) of two-mode quadrature observables and the logarithmic quantum Fisher informaiton $\ln(F_{\kappa}+1)$ as a function of dimensionless length for the parameters $\{\alpha=10$ and $\kappa=0.5\}$ and different $b$ values. Again, the solid lines represent $\ln\left(\Delta\kappa^{-2}_{d_1}+1\right)$ or $\ln\left(\Delta\kappa_{d_2}^{- 2}+1\right)$ while the dashed lines represent $\ln(F_{\kappa}+1)$. The insets depict the ratios of $\Delta\kappa^{-2}_{d_1}/F_{\kappa}$ and $\Delta\kappa^{-2}_{d_2}/F_{\kappa}$ (i.e., the Cram\'{e}r-Rao bound), respectively.}
\label{fig:two_mode_inverse_variances_F_k}
\end{figure}

\emph{\textcolor{blue}{Two-Mode Quadratures Case}}---Thanks to the simplicity of the system, we are able to carry out an analytical evaluation of quantum sensing performance by measuring two-mode quadratures as well. This further enables us to compare the results with those obtained from the single-mode quadrature case discussed right above. 

In order to have a quantitative understanding of the quantum sensitivity offered by two-mode quadrature detection, our objective is to find out the corresponding inverse variances and compare them with the quantum Fisher information $F_{\kappa}$, since the latter remains the same for the same system. In other words, we need to examine the Cram\'{e}r-Rao bound as implemented in the single-mode quadrature case. With use of Eqs.~(\ref{eq:d1var})-(\ref{eq:d2var}) and Eqs.~(\ref{eq:d1xk}) and (\ref{eq:d2xk}), we can easily derive the expressions for the inverse variances of the two-mode quadratures $d_1$ and $d_2$ as follows,
\begin{subequations}
\begin{align}
\Delta\kappa_{d_1}^{- 2} &= \frac{2\left(\chi^{d_1}_{\kappa}\right)^2}{\left\langle{\Delta q^2_i(0)} \right\rangle + \left\langle {\Delta q^2_s(l)} \right\rangle}=\frac{\left(\chi_{\kappa}^{q_{i}{(0)}}+\chi_{\kappa}^{q_{s}{(l)}}\right)^{2}}{\left\langle{\Delta q^2_i(0)} \right\rangle + \left\langle{\Delta q^2_s(l)} \right\rangle}, \label{eq:d1dk}\\
\Delta\kappa_{d_2}^{- 2} &= \frac{2\left(\chi^{d_2}_{\kappa}\right)^2}{\left\langle{\Delta p^2_i(0)} \right\rangle + \left\langle{\Delta p^2_s(l)} \right\rangle}=\frac{\left(\chi_{\kappa}^{p_{i}{(0)}}+\chi_{\kappa}^{p_{s}{(l)}}\right)^{2}}{\left\langle{\Delta p^2_i(0)} \right\rangle + \left\langle{\Delta p^2_s(l)} \right\rangle}.\label{eq:d2dk}
\end{align}
\end{subequations}
A quick comparison with the inverse variances (\ref{eq:qidk})--(\ref{eq:psdk}) in the single-mode quadrature case unveils a nontrivial constructive interference occurring in the two-mode quadrature scenario. This constructive interference is anticipated to provide better sensitivity than the single-mode quadrature case. To validate our hypothesis, in Fig.~\ref{fig:two_mode_inverse_variances_F_k}, we have compared the logarithmic $\Delta\kappa_{d_1}^{- 2}$ and $\Delta\kappa_{d_2}^{- 2}$ with the logarithmic $F_{\kappa}$. Further comparison with the single-mode quadrature case illustrated in Fig.~\ref{fig:inverse_variances_F_k} allows us to infer that both optimal classical sensing and (sub)optimal quantum sensing can be improved at present. To facilitate comprehension, we have also directly examined the ratios of $\Delta\kappa^{-2}_{d_1}/F_{\kappa}$ and $\Delta\kappa^{-2}_{d_2}/F_{\kappa}$ (i.e., the Cram\'{e}r-Rao bound) and represented them in the insets of Figs.~\ref{fig:two_mode_inverse_variances_F_k}(a) and (b) for your convenience. In these insets, a peak approaching one indicates best or optimal sensitivity, while other peaks with values less than one correspond to suboptimal sensitivities. However, whether these sensitivities are classical or quantum in nature depends on the characteristics of the observable involved at these peak locations.

In fact, this expectation can be supported by a mathematical proof. In the following, we outline such a proof. $\blacksquare$

For $\chi_{\kappa}^{j}$ and $\langle\Delta^2_j\rangle$ with $j\in\{q_i(0),p_i(0),q_s(l),p_s(l)\}$, we readily have the following inequalities:
\begin{subequations}
\begin{align}
\frac{{(\chi_{\kappa}^i)^2}\langle {\Delta^2_s}\rangle}{\langle {\Delta^2_i}\rangle}+\frac{{(\chi_{\kappa}^{s})^2}\langle {\Delta^2_i}\rangle}{\langle {\Delta^2_s}\rangle}&\geq2\chi_{\kappa}^{i}(\chi_{\kappa}^{s}),\nonumber\\
\frac{{(\chi_{\kappa}^{i})^2}(\left\langle {\Delta^2_s} \right\rangle+\left\langle {\Delta^2_i} \right\rangle)}{\left\langle {\Delta^2_i} \right\rangle}+\frac{{(\chi_{\kappa}^{s})^2}(\left\langle {\Delta^2_s} \right\rangle+\left\langle {\Delta^2_i} \right\rangle)}{\left\langle {\Delta^2_s} \right\rangle}&\geq{(\chi_{\kappa}^{i}+\chi_{\kappa}^{s})^2},\nonumber\\
\frac{(\chi_{\kappa}^{i})^2}{\left\langle {\Delta^2_i} \right\rangle}+\frac{(\chi_{\kappa}^{s})^2}{\left\langle {\Delta^2_s} \right\rangle}&\geq\frac{(\chi_{\kappa}^{i}+\chi_{\kappa}^{s})^2}{\left\langle{\Delta^2_i} \right\rangle+\left\langle{\Delta^2_s} \right\rangle}.\nonumber
\end{align}
\end{subequations}
In terms of the inverse variances, the last inequality can be rewritten as 
\begin{eqnarray}
\frac{(\chi_{\kappa}^{i}+\chi_{\kappa}^{s})^2}{\left\langle{\Delta^2_s} \right\rangle+\left\langle{\Delta^2_i} \right\rangle}\leq{\Delta\kappa_i^{- 2}+\Delta\kappa_s^{- 2}}. \label{eq:inference1}
\end{eqnarray}
Now with the additional help of the first and second inequalities, we can show that
\begin{subequations}
\begin{align}
\Delta\kappa_{d_1}^{- 2} &= \frac{\left(\chi_{\kappa}^{q_{i}{(0)}}+\chi_{\kappa}^{q_{s}{(l)}}\right)^{2}}{\left\langle{\Delta q^2_i(0)} \right\rangle + \left\langle{\Delta q^2_s(l)} \right\rangle}\leq{\Delta\kappa_{q_i(0)}^{- 2}+\Delta\kappa_{q_s(l)}^{- 2}}, \label{eq:d1dk_inequality}\\
\Delta\kappa_{d_2}^{- 2} &= \frac{\left(\chi_{\kappa}^{p_{i}{(0)}}+\chi_{\kappa}^{p_{s}{(l)}}\right)^{2}}{\left\langle{\Delta p^2_i(0)} \right\rangle + \left\langle{\Delta p^2_s(l)} \right\rangle}\leq{\Delta\kappa_{p_i(0)}^{- 2}+\Delta\kappa_{p_s(l)}^{- 2}}. \label{eq:d2dk_inequality}
\end{align}
\end{subequations}
The conditions under which the equality sign in Eqs.~(\ref{eq:d1dk_inequality}) and (\ref{eq:d2dk_inequality}) holds are, respectively, if and only if $\frac{\chi_{\kappa}^{q_{i}{(0)}}}{\chi_{\kappa}^{q_{s}{(l)}}}\sqrt{\frac{\left\langle{\Delta q^2_s(l)} \right\rangle}{\left\langle{\Delta q^2_i(0)} \right\rangle}}=\sqrt{\frac{\left\langle{\Delta q^2_i(0)} \right\rangle}{\left\langle{\Delta q^2_s(l)} \right\rangle}}$ and $\frac{\chi_{\kappa}^{p_{i}{(0)}}}{\chi_{\kappa}^{p_{s}{(l)}}}\sqrt{\frac{\left\langle{\Delta p^2_s(l)} \right\rangle}{\left\langle{\Delta p^2_i(0)} \right\rangle}}=\sqrt{\frac{\left\langle{\Delta p^2_i(0)} \right\rangle}{\left\langle{\Delta p^2_s(l)} \right\rangle}}$. 
The physics underlying these inequalities (\ref{eq:d1dk_inequality}) and (\ref{eq:d2dk_inequality}) suggests that, in the scenario of two-mode quadratures, apart from the upper limit imposed by the quantum Fisher information $F_\kappa$, there exists an additional lower limit representing the theoretically optimal boundary for measurement accuracy. To elaborate further, through mathematical analysis, the condition for achieving optimal and maximum quantum measurement accuracy for $\Delta\kappa_{d_1}^{- 2}$ is found to be when $nT$ satisfies the condition $4n\pi\textrm{Exp}\left[-\frac{ng\pi}{\beta}\right]=\frac{\beta}{\kappa}$, where the parameters $\{g,\kappa\}$ are involved. $\blacksquare$

Similar to the single-mode quadrature case, we do not observe any sensitivity improvement in the vicinity of the EP even with use of $d_1$. Moreover, the only area where enhanced sensitivity becomes approachable is within the PT-phase unbroken region, away from the EP. Furthermore, upon comparing the insets in Fig.~4 of the main text with those in Fig.~\ref{fig:two_mode_inverse_variances_F_k}, it becomes obvious that measuring two-mode quadratures can get better sensitivity than measuring single-mode quadratures, whether in classical or quantum contexts.

In summary, as one can see from the above derivations and discussions, it becomes evident that measuring a two-mode quadrature offers superior sensitivity compared to measuring a single-mode quadrature in type-II quadrature PT symmetry. Most importantly, our discoveries emphasize that when dealing with the quantum Fisher information $F_{\kappa}$ of a given quantum system, our selection of a suitable observable plays an essential role in determining how close we can approach this limit. Once again, it's worth pointing out that different observables may result in varying sensitivity and signal-to-noise ratio (SNR). Exploring the sensing performance offered by relative-intensity squeezing measurements in comparison to single-mode and two-mode quadrature cases could prove insightful. However, we defer this investigation to future research endeavors.